\documentclass[useAMS,usenatbib,usegraphicx]{mn2e}
\usepackage{times}
\newif\ifAMStwofonts
\AMStwofontstrue
\newcommand{\simlt}{\lower.5ex\hbox{$\; \buildrel < \over \sim \;$}}
\newcommand{\simgt}{\lower.5ex\hbox{$\; \buildrel > \over \sim \;$}}
\newcommand{\be}{\begin{equation}}
\newcommand{\ba}{\begin{eqnarray}}
\newcommand{\ee}{\end{equation}}
\newcommand{\ea}{\end{eqnarray}}

\title[Beyond SSPs]
{Exploring the Star Formation History of Elliptical Galaxies:
Beyond Simple Stellar Populations with a New Line Strength Estimator}
\author[B. Rogers et al.] {Ben Rogers$^1$, 
Ignacio Ferreras$^{1,2}$\thanks{E-mail: ferreras@star.ucl.ac.uk},
Reynier Peletier$^3$, Joseph Silk$^4$\\
$^1$ Department of Physics, King's College London, Strand, London WC2R 6LS\\
$^2$ Mullard Space Science Laboratory, Department of Space and Climate Physics, 
University College London, Holmbury St Mary, Dorking, Surrey RH5 6NT\\
$^3$ Kapteyn Astronomical Institute, University of Groningen, Postbus 800, 
9700 AV Groningen, Netherlands\\
$^4$ Physics Department, The Denys Wilkinson Building, Keble Road, Oxford, OX1 3RH
}

\begin{document}
\date{MNRAS, Accepted 2009 October 16}
\pagerange{\pageref{firstpage}--\pageref{lastpage}} \pubyear{2009}

\maketitle
\label{firstpage}

\begin{abstract}
We study the stellar populations of a sample of 14 elliptical galaxies
in the Virgo cluster. Using spectra with high signal-to-noise ratio
(S/N$\simgt 100$\AA\ $^{-1}$) we propose an alternative approach to
the standard side-band method to measure equivalent widths (EWs).  Our
\emph{Boosted Median Continuum} is shown to map the EWs more robustly
than the side-band method, minimising the effect from neighbouring
absorption lines and reducing the uncertainty at a given signal to
noise ratio. Our newly defined line strengths are more sucessful at
disentangling the age-metallicity degeneracy. We concentrate on
Balmer lines (H$\beta$,H$\gamma$,H$\delta$), the G band and the
combination [MgFe] as the main age and metallicity indicators. We go
beyond the standard comparison of the observations with simple stellar
populations (SSP) and consider four different models to describe the
star formation histories, either with a continuous star formation rate
or with a mixture of two different SSPs. These models improve the
estimates of the more physically meaningful mass-weighted
ages. Composite models are found to give more consistent fits among
individual line strengths and agree with an independent estimate using
the spectral energy distribution.  A combination of age and
metallicity-sensitive spectral features allows us to constrain the
average age and metallicity. For a Virgo sample of elliptical galaxies
our age and metallicity estimates correlate well with stellar mass or
velocity dispersion, with a significant threshold around $5\times
10^{10} M_\odot$ above which galaxies are uniformly old and metal
rich. This threshold is reminiscent of the one found by Kauffmann et
al. in the general population of SDSS galaxies at a stellar mass
$3\times 10^{10}M_\odot$. In a more speculative way, our models
suggest that it is formation \emph{epoch} and not formation timescale
what drives the Mass-Age relationship of elliptical galaxies.
\end{abstract}


\section{Introduction}

Unveiling the star formation histories of elliptical galaxies is key
to our understanding of galaxy formation. Being able to resolve their
seemingly homogenous distribution is hampered by the fact that their
light is dominated by old, i.e. low-mass stars, which do not evolve
significantly even over cosmological times. Furthermore, the presence
of small amounts of young stars as recently discovered in NUV studies
\citep{fs00,yi05,kav07} reveals a complex history of star formation
that requires proper estimates of mass-weighted ages, in contrast with
the luminosity-weighted ages that simple stellar populations (SSPs)
can only achieve. The majority of papers dealing with age estimates of
the stellar populations of elliptical galaxies rely on such SSPs
\citep[see e.g. ][]{kd98,sct00,tm05}, and it is only recently that special
emphasis has been made on the need to go beyond simple populations
\citep{fyi04,serra07,idi07}

Dating the (old) stellar populations of elliptical galaxies has been
fraught with difficulties, the most prominent being the
age-metallicity degeneracy, whereby the photo-spectroscopic properties
of a galaxy of a given age and metallicity can be replicated by a
younger or older galaxy at a suitably higher or lower metallicity,
respectively. To some extent, this problem has been overcome by the
measurement of pairs of absorption line indices \citep{wo94}, one
whose change in equivalent width (EW) is dominated by the average
metallicity of the population and the other dominated by the average
age of the population. Typical metal-sensitive line strengths are the
Mg feature at 5170\AA\ , the iron lines around 5300\AA , or a
combination such as [MgFe] \citep{gon93}. Balmer lines are more
age-sensitive and are often combined with metal-sensitive lines to break the
degeneracy. However, measurements of EWs of Balmer lines can be
affected by the age-metallicity degeneracy because of the presence of
nearby absorption lines. Such is the case of H$\gamma$, with the
prominent G band at 4300\AA\ or the CN bands in the vicinity of the
H$\delta$ line. This paper is partly motivated by the need to define a
method to estimate EWs that minimise the sensitivity of metallicity on
Balmer lines by a proper estimate of the continuum.

Even at relatively younger ages (a few Gyr) where such uncertainties
are reduced, considerable degeneracies still remain.  The confirmation
of recent star formation (RSF) occuring in early type galaxies
\citep{yi05,kav07} has raised the problem that the existence of a
young population can considerably affect the parameters derived
through SSP analysis. \citet{sct00} and later \citet{serra07} showed
that even relatively small mass fractions ($\sim$1\%) of young stars 
can distort age and metallicity estimates and in moderate cases
($\sim$10\%) completely overshadow the older population.  In addition,
the age and mass fraction of any younger sub-population will also be
degenerate, with larger mass fraction of relatively older sub
populations having the same effect as smaller fractions of younger
ones.

\citet{schiavon04} noticed that using different Balmer lines
(H$\beta$, H$\gamma$ and H$\delta$) to estimate the age gives slightly
different results, which was suggested to show that the galaxy had
undergone recent star formation.  Contrary to this, \citet{tm04} find
that H$\gamma$ and H$\delta$ equivalent widths are more affected by a
non-solar $\alpha$/Fe ratio on higher order Balmer lines.  This effect
is due to the increase of metal lines at bluer wavelengths, thereby
distorting the continuum as measured by a side-band method.  This was
expanded by \cite{serra07} who remodelled synthetic 2-burst models
using H$\beta$ and H$\gamma_A$ and achieved a result consistent with
both papers, concluding that a mistmatch between the three Balmer line
estimates could possibly reveal underlying younger populations.

Moving forward with such analysis -- beyond simple populations and
luminosity-weighted parameters -- requires improvements on both the
H$\gamma$ and H$\delta$ measurements. Balmer line equivalent widths
suffer from the effects of the metallicity due to the presence of such
lines in the spectral region used to determine the continuum
\citep{wo97,tm04,pro07}. Although considerable work has already been
done in this area \citep[see e.g. ][]{rose94,jw95,va99,yam06}, we test
a completely different approach.

In this paper we present a comprehensive analysis of the stellar
populations of 14 Virgo cluster ellipticals using several models to
describe the star formation history, exploring the discrepancies found
between simple and composite stellar populations. In an attempt to
combat the problems discussed above, we introduce a new method for the
measurement of equivalent widths using a high percentile running
``median'' to describe the continuum.  The properties of this new method
are exploited by using various age- and metallicity-sensitive spectral
features.

\begin{figure*}
\begin{minipage}{18cm}
\begin{center}
\includegraphics[width=3.3in]{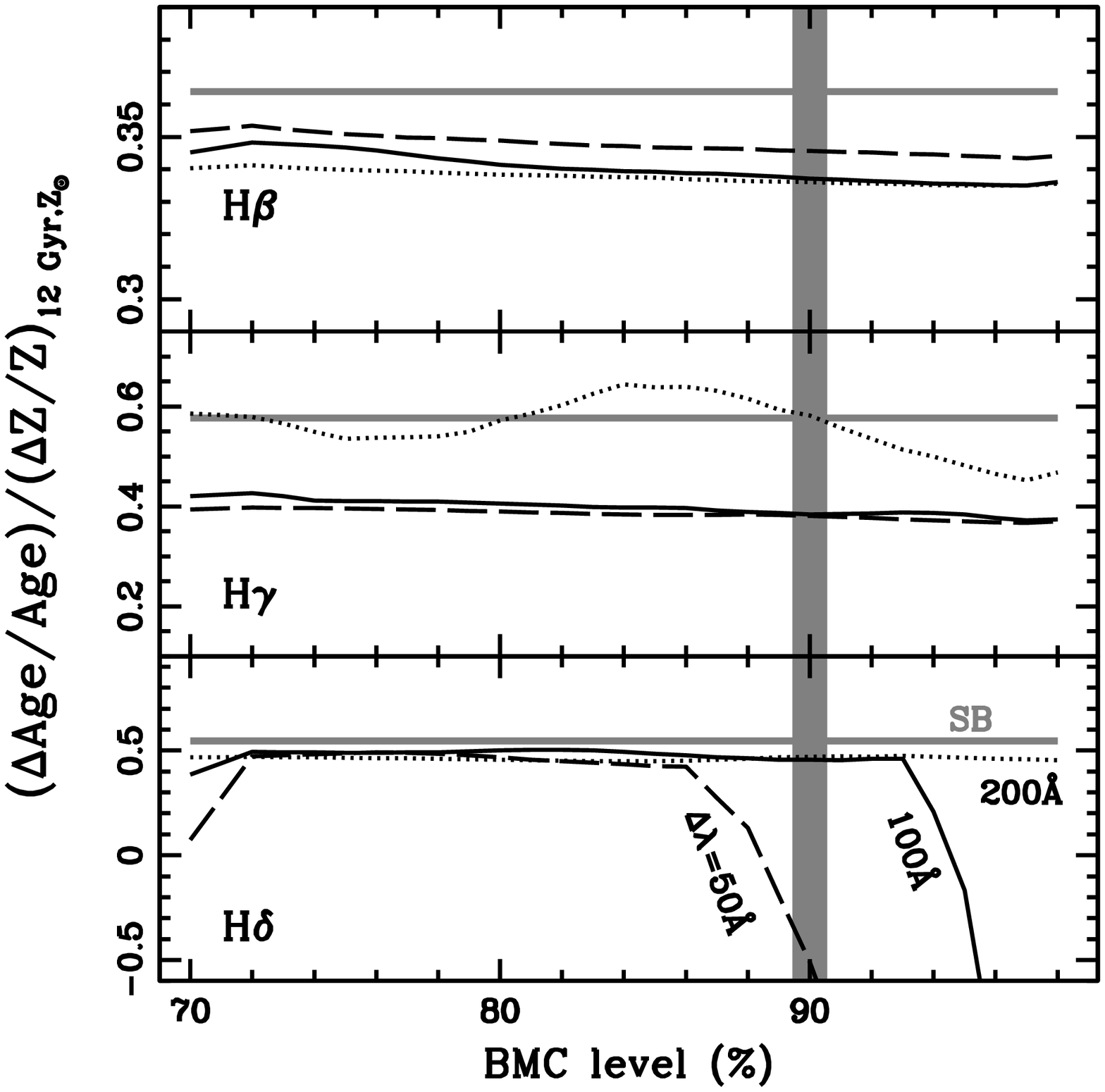}
\includegraphics[width=3.3in]{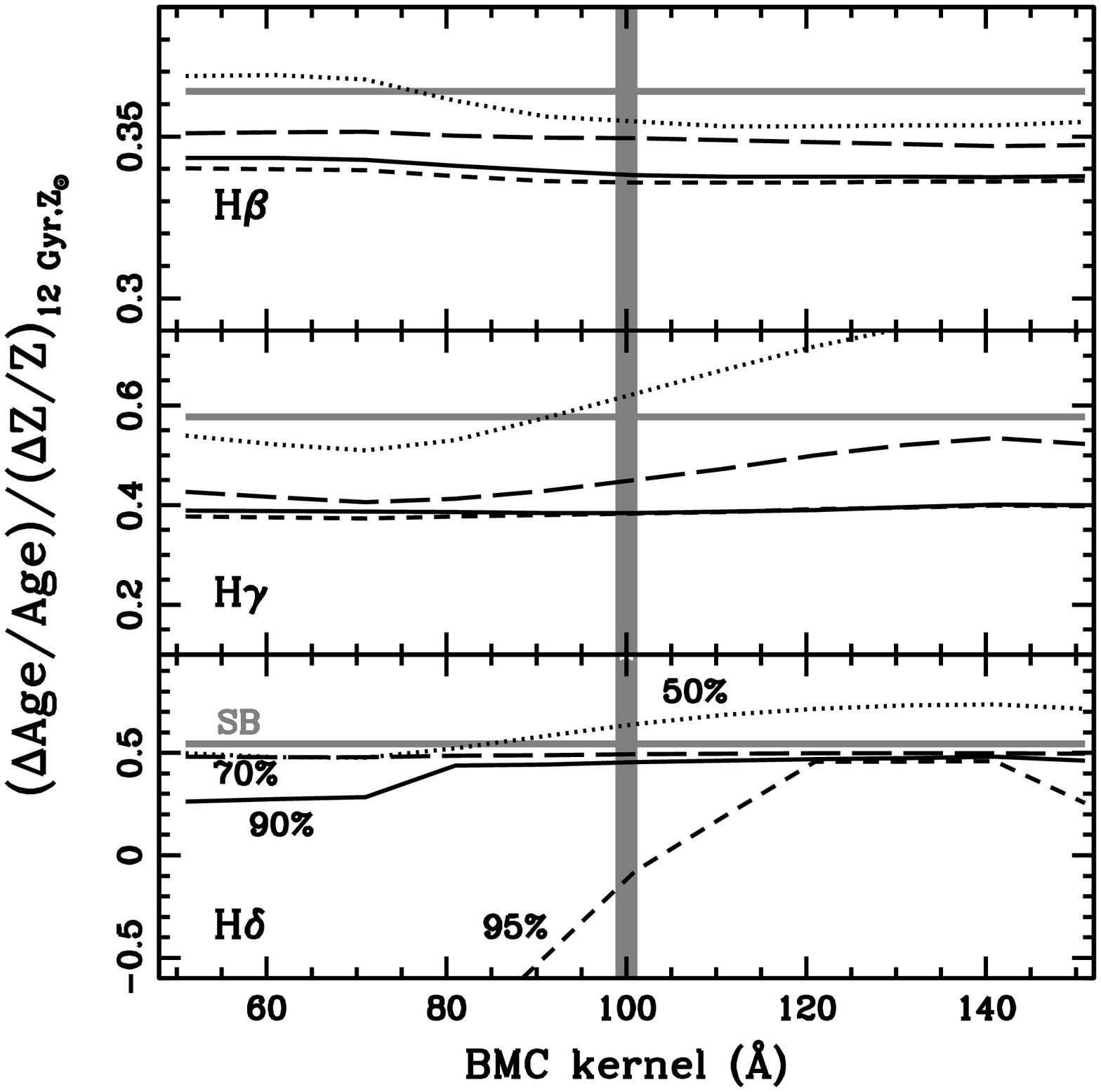}
\caption{Dependence of the age-metallicity gradient -- ($\Delta$
Age/Age)(/$\Delta$ Z/Z) -- on the choice of BMC parameters, namely the
level at which the 'boosted median' is taken (\emph{left}) or the size
of the kernel (\emph{right}). On the left panel a number of kernel
sizes is shown as labelled: $\Delta\lambda$=50\AA\ (dashed); 100\AA\
(solid) and 200\AA\ (dotted). On the right panel a number of
confidence levels are considered: 50\% (dotted); 70\% (long dashed);
90\% (solid) and 95\% (short dashed).  The grey horizontal line in
both panels is the estimate from the standard side-band method (SB), and the
grey vertical shaded area marks our choice of 'boosted median' parameters.}
\label{fig:DADZ}
\end{center}
\end{minipage}
\end{figure*}

\section{The sample}

We use a sample of 14 elliptical galaxies in the Virgo cluster, for
which moderate resolution spectroscopy is available at high
signal-to-noise ratio \citep[S/N\simgt 100
\AA$^{-1}$,][]{yam06,yam08}.  Eight galaxies were observed with FOCAS
at the 8m Subaru telescope; the other six were observed with ISIS at
the 4.2m William Herschel Telescope (WHT). Observations from Subaru
span the spectral range $\lambda\simeq 3800-5800$\AA\ , whereas the
spectra taken at the WHT span a narrower window, namely $\lambda\simeq
4000-5500$\AA\ .  The resolution (FWHM) of both data sets is similar:
2\AA\ (Subaru) and 2.4\AA  (WHT). 
 We refer the interested reader to \citet{yam06} for
details about the data reduction process.  We compare those spectra
with composites of the R$\sim 2000$ synthetic models of
\citet{bc03}, updated to the 2007 version \citep{cb07}. 
We resampled the observed spectra from the original
0.3\AA\ to 1\AA\ per pixel, performing an average of the spectra over
a 1.5\AA\ window, in order to have a sampling more consistent with the
actual resolution.

In this paper we use two alternative sets of information, either the full
spectral energy distribution or targeted absorption lines. For the
former, we consider a spectral window
around the 4000\AA\ break, which is a strong age indicator (albeit
with a significant degeneracy in metallicity, especially for
evolved stellar populations). In order to minimise the effect of
an error in the flux calibration, we do not choose the full spectral
range of the spectra, restricting the analysis to 3800--4500\AA\  for the
Subaru spectra, and 4000--4500\AA\ for the WHT spectra.

The second method focuses on a reduced number of spectral
lines. Following the traditional approach \citep[see
e.g.][]{kd98,sct00,tm05,sbla06}, we use a set of age-sensitive and
metallicity-sensitive lines. In the next section we describe in detail
the indices targeted by our analysis and describe a new algorithm that
improves on the ``standard'' method to determine the continuum in
galaxy spectra.

\section{Measuring Equivalent Widths}

We focus on a reduced set of absorption lines originally defined in
the Lick/IDS system \citep{lick} and extensions thereof
\citep{wo97}. As age-sensitive lines we use the Balmer lines H$\beta$,
H$\gamma$, H$\delta$, the G-band (G4300) and the 4000\AA\ break
(D4000).  We use the standard definition of [MgFe] \citep{gon93}, as a
metal-sensitive tracer, which is a reliable proxy of overall
metallicity, with a very mild dependence on [$\alpha$/Fe] abundance
ratio \citep{tm03}.

The standard method to determine the equivalent widths of galaxy
spectra relies on the definition of a blue and a red side-band to
determine the continuum. A linear fit to the average flux in the blue 
and red side-bands is used to track the continuum in the line
\citep[see e.g.][]{sct00}. This method, although easy to implement,
has an important drawback as neighbouring lines can make a significant
contribution to the flux in the blue and red passbands, introducing
unwanted age/metallicity effects. For instance, the H$\gamma$ index
\citep{wo97} is defined with the blue side-band located close to the
prominent G-band, around 4300\AA. This definition causes
non-physical negative values of the H$\gamma$ line in \emph{absorption}, 
as the depression caused by the G-band makes the flux in
the $H\gamma$ line (wrongly) appear in \emph{emission}. This has not
prevented the community from using this line as a sensitive age tracer, as
long as models and data are treated in the same way. \cite{va99}
defined new measurements of this line in order to reduce the
metallicity degeneracy mainly introduced by the choice of the side
bands.  They avoid this by selecting specific regions
less affected by the metal absorption lines.

H$\delta$ is another Balmer line which has been recently considered to
be affected by neighbouring metal lines -- most notably the CN molecular bands --
which reduce the age sensitivity of the index \citep{pro07}.

In this paper we present an alternative method to determine the equivalent
widths. Our method does not rely on the definition of blue and red side-bands
and minimizes the contamination from neighbouring lines. This 
method is simple to apply and we propose it for future studies of stellar
populations in galaxies\footnote{A C-programme that computes EWs using our
proposed BMC method from an ASCII version of an SED can be obtained from
us (\emph{ferreras@star.ucl.ac.uk}).}.

\subsection{The Boosted Median Continuum (BMC)}

Our measure of equivalent width follows the standard procedure
comparing observed flux in the line and the corresponding
``interpolated'' continuum in the same wavelength range. For an
equivalent width measured in \AA:

\be
EW = \int_{\lambda_1}^{\lambda_2} \Big[ 1-\frac{\Phi(\lambda)}{\Phi_C(\lambda)}\Big]
d\lambda,
\ee 

\noindent
where $\lambda_1$ and $\lambda_2$ define the wavelength range of the
line, $\Phi(\lambda)$ is the observed flux, and $\Phi_C(\lambda)$ is
the flux from the continuum. Rather than defining the continuum as a
linear fit between a blue and a red side-band, we propose the
``boosted median'' of the flux, defined at each wavelength as the 90th
percentile of the flux values within a 100\AA\ window.

This method is defined by two parameters, namely the choice of
percentile (90\% in our case) and the size of the kernel
($\Delta\lambda=$100\AA).  The kernel size needs to be large enough to
avoid the small scale variations of the spectra, but small enough to
avoid distortions from large scale structure of the spectra such as
the breaks at 4000\AA\ and 4300\AA, or flux calibration errors
\footnote{Applying a flux calibration distortion of ~5\% --
consistent with that found in current surveys (SDSS) -- 
causes negligible effects on the EW measured with the BMC method (below 1\%).}.
The choice of percentile also suffers a similar balancing act, since
it should be high enough to select the true continuum but at a value
that would avoid it becoming dominated by noise. In order to determine
the optimal choice, a range of values for these two parameters was
studied on a number of simple stellar populations taken from the
models of \citet{bc03} including the effect of velocity dispersion and
noise.  Out of the simulations, we adopted the 90th percentile of the
flux within a 100\AA\ window. However, this choice is not critical,
given the robustness of the method in which the continuum is selected
(i.e. a median). The advantage of averaging over a large enough
wavelength range is that the effect of strong metallic lines in the
vicinity of the index is limited. Furthermore, this pseudo-continuum
is found to be less susceptible to noise (see below).

\begin{table*}
\caption{Equivalent Widths of Virgo Elliptical galaxies using a 90\% 
Boosted Median Continuum (see text for details). 
All values given in \AA\ and measured at the observed $\sigma$, with the 1$\sigma$
uncertanities in brackets below each measurement.}
\label{tab:EWs}
\begin{tabular}{lrccccccccc}
\hline\hline
Galaxy & $\sigma^1$ & H$\beta_{20}$ & H$\gamma_{20}$ & H$\delta_{20}$ & 
Mgb$_{20}$ & Fe5270$_{20}$ & Fe5335$_{20}$ & G4300$_{20}$ & D4000 & [MgFe]$_{20}$\\
\hline
NGC 4239 &  82 & 2.993  & 2.097  & 2.356  & 2.972  & 2.864  & 1.673  & 6.130  & 1.441  & 2.596 \\
 & & (0.051) & (0.064) & (0.056) & (0.044) & (0.046) & (0.060) & (0.046) & (0.006) & (0.032) \\
NGC 4339 & 142 & 2.666  & 1.488  & 1.442  & 3.807  & 3.163  & 1.772  & 6.642  & 1.560  & 3.065 \\
 & & (0.047) & (0.054) & (0.082) & (0.045) & (0.037) & (0.050) & (0.042) & (0.005) & (0.024) \\
NGC 4365 & 245 & 2.191  & 0.732  & 1.195  & 3.819  & 2.971  & 1.615  & 6.289  & ---  & 2.959 \\
 & & (0.026) & (0.042) & (0.070) & (0.023) & (0.028) & (0.031) & (0.021) & (---) & (0.015) \\
NGC 4387 & 105 & 2.431  & 1.698  & 1.974  & 3.614  & 3.237  & 1.703  & 7.317  & ---  & 2.988 \\
 & & (0.044) & (0.066) & (0.068) & (0.039) & (0.057) & (0.061) & (0.050) & (---) & (0.024) \\
NGC 4458 & 104 & 2.464  & 1.622  & 1.676  & 3.520  & 2.719  & 1.670  & 6.570  & 1.504  & 2.779 \\
 & & (0.039) & (0.053) & (0.055) & (0.037) & (0.042) & (0.047) & (0.041) & (0.006) & (0.025) \\
NGC 4464 & 135 & 2.302  & 1.547  & 1.523  & 3.648  & 2.913  & 1.521  & 6.813  & ---  & 2.844 \\
 & & (0.038) & (0.053) & (0.070) & (0.035) & (0.034) & (0.036) & (0.037) & (---) & (0.023) \\
NGC 4467 &  75 & 2.605  & 1.797  & 1.875  & 3.839  & 3.209  & 1.747  & 6.760  & 1.458  & 3.085 \\
 & & (0.050) & (0.063) & (0.065) & (0.044) & (0.052) & (0.056) & (0.049) & (0.005) & (0.026) \\
NGC 4472 & 306 & 2.225  & 0.793  & 1.149  & 3.650  & 2.719  & 1.535  & 5.742  & 1.525  & 2.787 \\
 & & (0.031) & (0.069) & (0.081) & (0.039) & (0.039) & (0.042) & (0.027) & (0.005) & (0.021) \\
NGC 4473 & 180 & 2.410  & 1.233  & 1.592  & 3.990  & 3.212  & 1.808  & 6.918  & ---  & 3.164 \\
 & & (0.028) & (0.044) & (0.060) & (0.026) & (0.023) & (0.041) & (0.020) & (---) & (0.021) \\
NGC 4478 & 132 & 2.627  & 1.653  & 1.855  & 3.719  & 3.378  & 1.838  & 6.764  & ---  & 3.114 \\
 & & (0.042) & (0.051) & (0.069) & (0.036) & (0.039) & (0.040) & (0.039) & (---) & (0.023) \\
NGC 4489 &  73 & 3.204  & 2.084  & 2.328  & 3.217  & 3.248  & 1.903  & 6.455  & 1.520  & 2.878 \\
 & & (0.047) & (0.054) & (0.075) & (0.046) & (0.048) & (0.045) & (0.050) & (0.005) & (0.028) \\
NGC 4551 & 105 & 2.663  & 1.717  & 1.736  & 3.907  & 3.416  & 2.027  & 6.828  & 1.531  & 3.261 \\
 & & (0.046) & (0.058) & (0.056) & (0.034) & (0.038) & (0.044) & (0.036) & (0.005) & (0.024) \\
NGC 4621 & 230 & 2.284  & 0.964  & 1.312  & 4.072  & 3.014  & 1.678  & 6.250  & ---  & 3.091 \\
 & & (0.016) & (0.024) & (0.040) & (0.012) & (0.017) & (0.018) & (0.011) & (---) & (0.009) \\
NGC 4697 & 168 & 2.397  & 1.289  & 1.661  & 3.806  & 3.198  & 1.832  & 6.603  & 1.524  & 3.094 \\
 & & (0.029) & (0.037) & (0.071) & (0.027) & (0.026) & (0.033) & (0.021) & (0.004) & (0.017) \\
\hline
\end{tabular}

{$^1$ Velocity dispersions given in km/s, from \citet{yam06}.\hfill}
\end{table*}

\begin{figure}
\begin{center}
\includegraphics[width=3.2in]{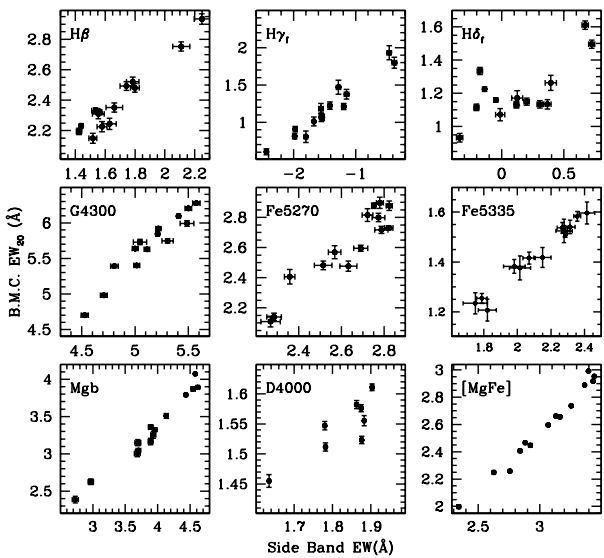}
\caption{Comparison between the EWs measured by the side-band method 
(SB; horizontal)
and our proposed 'Boosted Median Continuum' (BMC; vertical) for H$\beta$,
H$\gamma_F$, H$\delta_F$, G4300, Fe5270, Fe5335, Mgb, D4000 and
[MgFe]. The error bars are shown at the 1$\sigma$ level. Notice only 8
galaxies (observed with Subaru) have a measurement of the 4000\AA\
break. For the remaining six galaxies our data does not extend bluer
than 4000\AA .}
\label{fig:EWSBBMC}
\end{center}
\end{figure}

Figure~\ref{fig:DADZ} motivates our choice of parameters. We show the
age-metallicity sensitivity -- ($\Delta$Age/Age)/($\Delta$Z/Z) -- for
simple stellar populations measured at 12~Gyr and solar
metallicity. We do not follow the same definition as in
\citet{wo94}. Instead, we change the age from the reference value by
2~Gyr and find the change in metallicity required from the fiducial SSP
that gives the same variation in the EW.  The grey horizontal bar is
the value determined from the standard side-band method. Smaller
values of the gradient imply a better disentanglement of the
age-metallicity degeneracy.  In the left (right) panels the horizontal
axis explores a range of confidence levels (kernel sizes). The lines
correspond to various choices of kernel size (left) and confidence
level (right) as labelled.  H$\beta$ (\emph{top}) behaves quite
robustly with respect to the choice of BMC parameters. H$\gamma$
(\emph{middle}) shows that the presence of large scale features such
as the break found around the G band at 4300\AA\ can affect the
estimate if a large kernel size is chosen (200\AA, dotted line,
\emph{left}). Finally, H$\delta$ (\emph{bottom}) shows that too small
a kernel size ($\Delta\lambda=$50\AA, dashed line, \emph{ left}) or
too high a confidence level (95\%, short dashed line, \emph{ right})
can affect the estimate. In this case the effect is caused by nearby
features that will contaminate the estimate of the pseudo-continuum.
Our choice of 100\AA\ kernel size and 90\% level is thereby justified
by the need to avoid both short and long-scale features in the
SEDs. Furthermore, lower confidence levels should be avoided as they
will give flux values closer to an average that will not reflect the
true continuum given the presence of numerous absorption lines. Too
high values of the level will make the measurement more prone to
higher uncertainties at low signal-to-noise ratios.  Monte Carlo
simulations of noise show that for our choice of BMC parameters the
errors for the BMC EWs are always smaller than those from the
standard side-band method.

\begin{figure}
\begin{center}
\includegraphics[width=3.2in]{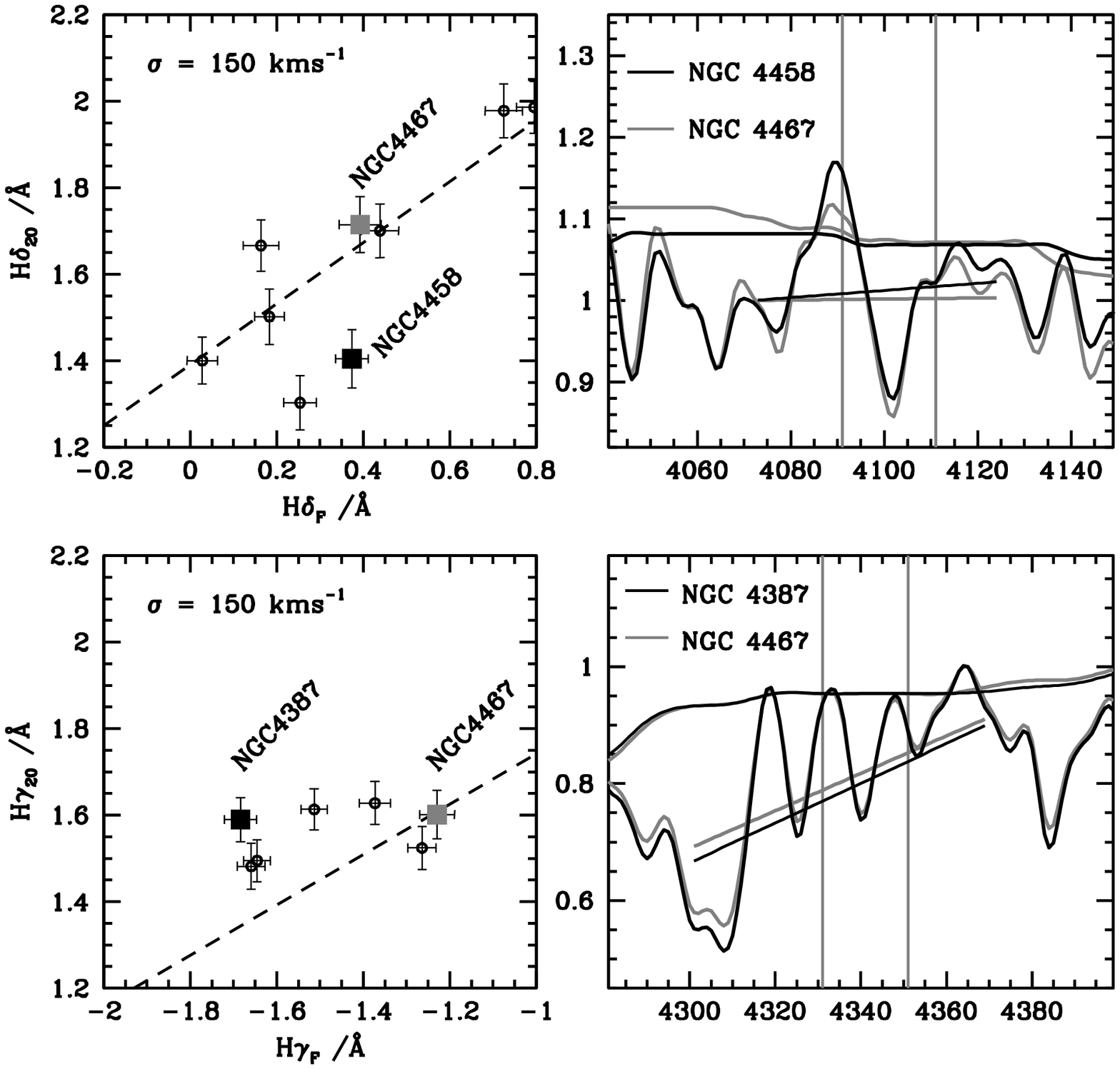}
\caption{Comparison of the H$\gamma_F$ and H$\delta_F$ equivalent
width measured with the side-band (SB) and the Boosted Median method
(BMC). 
In order to eliminate the dependence of the EWs on
velocity dispersion, we select those galaxies with $\sigma\leq 150$~km/s 
and smooth them to this maximum velocity dispersion.  \emph{Top}:
We select NGC 4458 \& NGC 4467, which show similar EWs in
the SB method but have differing H$\delta$ EWs when using BMC.  In the
rightmost panels we plot the spectra from both galaxies. Overplotted
are the pseudo-continuum of the side band (slanted straight line) and
the BMC (running along the top of the SED). The identical EWs
according to the SB method are due to the decreased flux in the red
passband of NGC 4467, most likely caused by increased CN absorption,
which hides the stronger intrinsic absorption in H$\delta$. The
spectra are normalised to an average value of 1 across the blue
passband of H$\delta_F$ to highlight the effect of CN absorption.
\emph{Bottom}: We highlight H$\gamma$ in NGC 4387 \& NGC 4467,
for which the SB method indicates a considerable difference in EW
while the BMC identifies almost none. Looking at the spectra (right) of
both galaxies we can see that the difference in EW from the SB method
is caused by an increase in the depth of the G4300 feature,
artificially lowering the pseudo-continuum.  The spectra are
normalised to 1 at 4365\AA .}
\label{fig:Balmer1}
\end{center}
\end{figure}

\begin{figure}
\begin{center}
\includegraphics[width=3.2in]{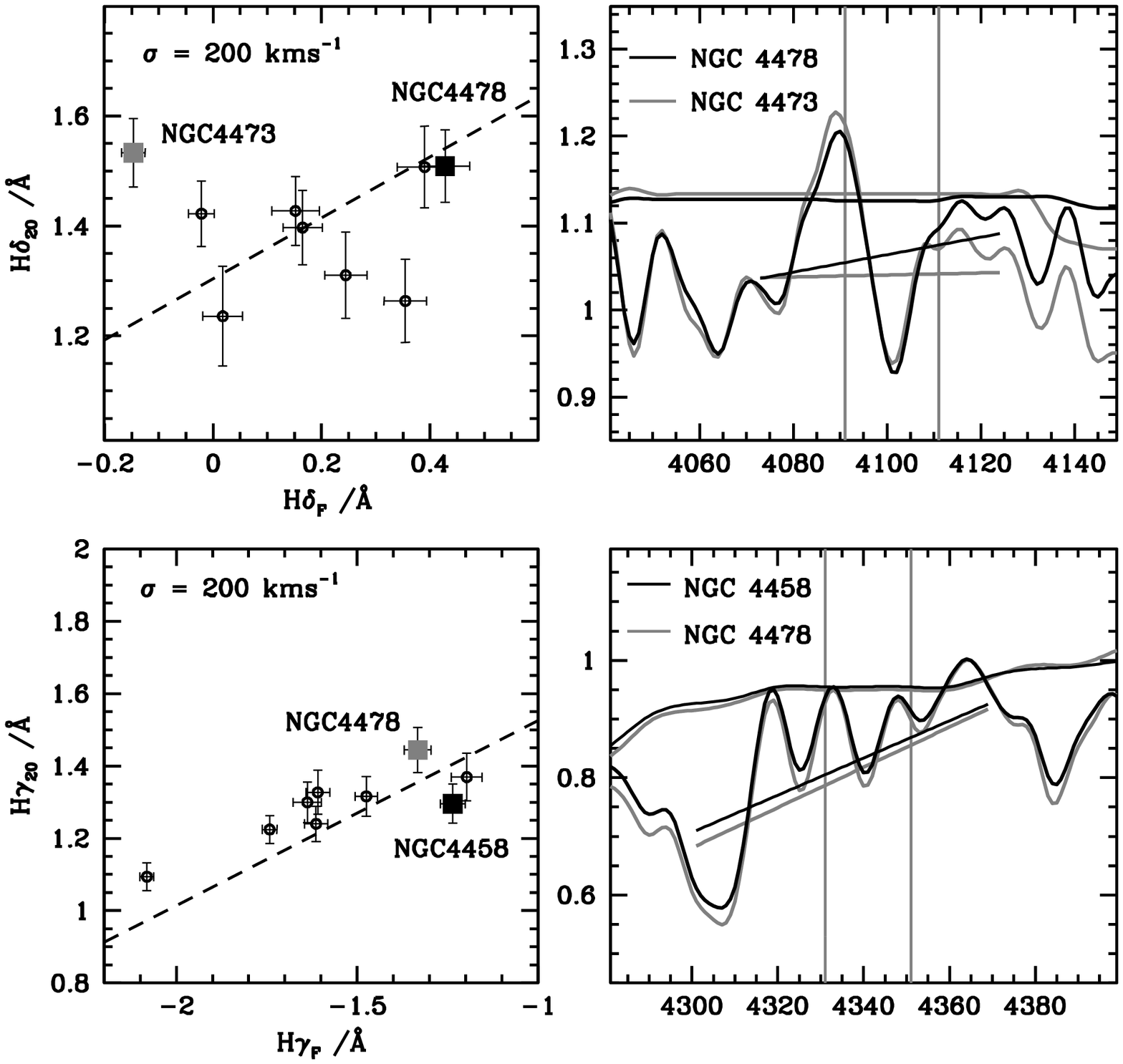}
\caption{As in figure~\ref{fig:Balmer1} we compare the H$\gamma_F$
and H$\delta_F$ equivalent widths measured with the side-band and the
Boosted Median method. In order to eliminate the dependence of the EWs
and spectra on velocity dispersion, we select -- in contrast to
figure~\ref{fig:Balmer1} -- those galaxies with $\sigma\leq 200$ km/s.
The SEDs are smoothed to this maximum velocity dispersion.
\emph{Top}: The galaxies NGC 4473 \& NGC 4478 are selected since they
show differing H$\delta$ SB EWs but similar BMC measured EWs.  The
spectra (right) identifies the cause of the discrepancy. The spectra
of NGC 4473 redwards of the H$\delta$ line is severely affected by CN
absorption, lowering the average flux in this side band and distorting
the derived pseudo-continuum. The BMC recovers a significantly more
robust value. The spectra are normalised to an average value of 1
across the blue passband of H$\delta_F$ to highlight the effect of CN
absorption.  \emph{Bottom}: NGC 4458 \& NGC 4478 show
discrepant SB EWs. This difference
can be seen on the right, where the spectra of both galaxies are
displayed. The BMC correctly identifies the larger value of H$\gamma$
in NGC 4478, which is hidden in the SB method due to the increased absorption
of G4300 feature located in the blue pass band of H$\gamma_F$. The
spectra are normalised to 1 at 4365\AA.}
\label{fig:Balmer2}
\end{center}
\end{figure}

Hence, the method for generating the BMC is fairly simple. At every
wavelength the 90th percentile from all flux measurements within
50\AA\ on either side is assigned as the continuum at that wavelength.
With the continuum thus defined, for each line strength we only have
to define the central wavelength and the spectral window over which
the line is measured. Given that our method maps quite robustly the
underlying continuum, we decided to fix a 20\AA\ width for all line
strengths considered in this paper -- hereafter, our BMC-based
equivalent widths are labelled with a $20$ subindex.

\begin{table*}
\caption{Range of parameters explored in this paper (see text
for a description of each model).}
\label{tab:params}
\begin{tabular}{lccl}
\hline\hline
Model/Param & MIN & MAX & Comments\\
\hline
SSP  & & & 2 params\\
Age  & 3 & 13 & Gyr\\
$\log(Z/Z_\odot)$ & -1.5 & +0.3 & Metallicity\\
\hline
EXP  & & & 3 params\\
$\log\tau$ (Gyr) & -1 & +0.9 & Exp. Timescale\\
$z_F$ & 0.1 & 5 & Formation epoch\\
$\log(Z/Z_\odot)$ & -1.5 & +0.3 & Metallicity\\
\hline
2BST & & & 4 params\\
t$_{\rm O}$ & 3   & 13  & Old (Gyr)\\
t$_{\rm Y}$ & 0.1 &  3  & Young (Gyr)\\
f$_{\rm Y}$ & 0.0 & 0.5 & Mass fraction\\
$\log(Z/Z_\odot)$ & -1.5 & +0.3 & Metallicity\\
\hline
CXP  & & & 3 params\\
$\log\tau_1$ (Gyr) & -2 & +0.5 & SF Timescale\\
$\log\tau_2$ (Gyr) & -2 & +0.5 & Enrichment Timescale\\
$z_F$ & 0.1 & 5 & Formation epoch\\
\hline
\end{tabular}
\end{table*}

This is motivated by the fact that for some definitions of the line
strengths, any 'contaminating' lines falling within the central
bandpass can have a stronger effect on the BMC method compared to the
standard side-band method. A clear example is found in the $H\beta$
line, where the effect of FeI(~4871\AA) which sits on the shoulder of
the Balmer line within the standard definition of the central passband
(width 28.75\AA\ ) clearly distorts the measurement of the equivalent
width. This is slightly 'compensated' in the side-band method through
the presence in the blue side-band of another significant Fe line at
4891\AA. This is clearly not the case for the BMC method, in which the
pseudo-continuum is effectively independent of the values of both Fe
absortion lines. Hence, in the BMC method it is desirable to choose a
central passband which only targets the line of interest. 
Nevertheless, the method is versatile enough to define wider central
bandpasses.

Thus, to avoid this problem we define a central passband wavelength as
narrow as possible within the usual spectral resolutions targeted in
unresolved stellar populations. We follow the 'F' type passbands as
used for the Balmer indices \citep{wo97}, defining a 20\AA\ window
centered on the line of interest for all indices
considered\footnote{Our BMC-based EWS are labelled with a '20'
subindex, e.g. H$\beta_{20}$}. In the case of the metal line strengths -- 
which usually involve clusters of lines -- the major feature is chosen
as the center of the new 20\AA\  index. In the case of Mgb$_{20}$, this
is centered between the MgI doublet at 5167/5172\AA. Fe5270$_{20}$ is
defined by the FeI line at 5270\AA\ and Fe5335$_{20}$ uses the FeI
line at 5328\AA. The central passband of the G-band is 4300\AA.  Such
a simple approach is nevertheless very versatile in its definition of
any line strength. The new $D4000$ break feature uses the definition
given by \citet{bal99}, with the difference that it is the ratio of
the continuum flux obtained by the BMC method within those wavelengths
that is used here.

Table~\ref{tab:EWs} shows the (BMC-measured) equivalent widths of the
Virgo elliptical galaxies targeted in this paper. The measurements are
obtained directly from the observed spectra, presented in
\citet{yam06} and have not been corrected with respect to velocity
dispersion. Analogously to the standard method, one could either
correct the observed EWs for the effect of velocity dispersion, or use
the observed EWs in the modelling. For the latter, one must then
measure the model EWs on spectra with the same resolution and velocity
dispersion as the targeted galaxy. We follow this approach in the
paper.  We note at this point that the effect of velocity
dispersion on the EWs of Balmer lines measured with the BMC method is
increased slightly since the method uses spectral information to
better define the pseudo-continuum, which is destroyed by an
increasing $\sigma$. This effect of course affects any other method which
measures line strengths from unresolved spectra
\citep[e.g.][]{va99}.  The numbers in brackets correspond to the
1-$\sigma$ uncertainty, obtained from a Monte Carlo simulation that
generates 500 realizations of each SED, adding noise corresponding to
the SNR of the observations. Notice that 6 of the galaxies do not have
a measured D4000 as they were observed over a spectral range that does
not include the blue passband used in the definition of D4000.

\renewcommand{\arraystretch}{1.1}
\begin{table*}
\begin{center}
\begin{minipage}{18cm}
\caption{Ages and metallicities of Virgo elliptical galaxies
according to the four models used in this paper. Error bars
quoted at the 90\% confidence level.}
\label{tab:Ages}
\begin{tabular}{lrrc|rrc|rrc|rrc}
\hline\hline
Galaxy   & \multicolumn{3}{c}{SSP} &\multicolumn{3}{c}{2BST} &\multicolumn{3}{c}{EXP} &\multicolumn{3}{c}{CXP}\\
NGC & Age(Gyr) & log(Z/Z$_\odot$) & $\chi_r^2$  & Age(Gyr) & log(Z/Z$_\odot$) & $\chi_r^2$ & Age(Gyr) & log(Z/Z$_\odot$) & $\chi_r^2$ & Age(Gyr) & log(Z/Z$_\odot$) & $\chi_r^2$ \\
\hline
4239 & $ 5.0_{ -0.6}^{ +0.6}$ &   $-0.13_{-0.07}^{+0.07}$ &  0.29 &
 $ 6.6_{ -1.7}^{ +2.6}$ &   $-0.14_{-0.08}^{+0.08}$ &  0.27 &
 $ 5.3_{ -0.8}^{ +1.0}$ &   $-0.14_{-0.07}^{+0.07}$ &  0.27 &
 $ 5.5_{ -0.9}^{ +0.6}$ &   $+0.06_{-0.07}^{+0.04}$ &  0.83\\

4339 & $ 8.2_{ -1.6}^{ +1.6}$ &   $+0.18_{-0.10}^{+0.06}$ &  1.60 &
 $ 8.5_{ -1.9}^{ +1.7}$ &   $+0.19_{-0.10}^{+0.06}$ &  1.58 &
 $ 8.8_{ -1.7}^{ +1.5}$ &   $+0.16_{-0.09}^{+0.07}$ &  1.61 &
 $ 9.6_{ -1.7}^{ +1.5}$ &   $+0.25_{-0.05}^{+0.03}$ &  1.23\\

4365 & $11.6_{ -0.6}^{ +1.0}$ &   $+0.26_{-0.04}^{+0.03}$ &  2.30 &
 $11.7_{ -0.6}^{ +0.9}$ &   $+0.26_{-0.04}^{+0.03}$ &  2.30 &
 $11.4_{ -0.4}^{ +0.4}$ &   $+0.27_{-0.02}^{+0.02}$ &  2.76 &
 $12.1_{ -0.3}^{ +0.1}$ &   $+0.30_{-0.02}^{-0.00}$ &  3.11\\

4387 & $12.5_{ -0.9}^{ +0.4}$ &   $-0.09_{-0.05}^{+0.03}$ &  2.36 &
 $12.4_{ -1.1}^{ +0.5}$ &   $-0.09_{-0.05}^{+0.03}$ &  2.36 &
 $11.4_{ -1.2}^{ +0.5}$ &   $-0.06_{-0.03}^{+0.10}$ &  2.96 &
 $11.4_{ -0.9}^{ +0.7}$ &   $+0.15_{-0.05}^{+0.02}$ &  3.15\\

4458 & $10.7_{ -0.8}^{ +0.9}$ &   $-0.22_{-0.04}^{+0.04}$ &  4.34 &
 $10.7_{ -0.9}^{ +1.0}$ &   $-0.22_{-0.04}^{+0.04}$ &  4.34 &
 $10.7_{ -0.7}^{ +0.7}$ &   $-0.21_{-0.05}^{+0.04}$ &  5.06 &
 $11.6_{ -1.1}^{ +0.8}$ &   $-0.01_{-0.07}^{+0.06}$ &  1.90\\

4464 & $12.7_{ -0.5}^{ +0.2}$ &   $-0.14_{-0.03}^{+0.02}$ &  3.87 &
 $12.7_{ -0.4}^{ +0.2}$ &   $-0.14_{-0.03}^{+0.02}$ &  3.87 &
 $11.9_{ -0.4}^{ +0.2}$ &   $-0.13_{-0.02}^{+0.02}$ &  4.85 &
 $12.0_{ -0.6}^{ +0.4}$ &   $+0.09_{-0.03}^{+0.01}$ &  3.68\\

4467 & $ 6.4_{ -1.1}^{ +1.3}$ &   $+0.16_{-0.07}^{+0.06}$ &  2.79 &
 $ 7.6_{ -1.8}^{ +2.6}$ &   $+0.19_{-0.07}^{+0.07}$ &  2.78 &
 $ 6.8_{ -1.3}^{ +1.7}$ &   $+0.15_{-0.09}^{+0.06}$ &  3.33 &
 $10.0_{ -1.9}^{ +1.8}$ &   $+0.21_{-0.06}^{+0.04}$ &  0.67\\

4472 &  $ 9.2_{ -0.5}^{ +0.6}$ &   $+0.27_{-0.02}^{+0.02}$ &  2.55 &
 $ 9.4_{ -0.5}^{ +0.4}$ &   $+0.27_{-0.02}^{+0.02}$ &  2.55 &
 $ 9.8_{ -0.7}^{ +0.8}$ &   $+0.27_{-0.02}^{+0.02}$ &  2.46 &
 $11.3_{ -0.8}^{ +0.6}$ &   $+0.29_{-0.03}^{+0.00}$ &  2.19\\

4473 & $12.5_{ -0.8}^{ +0.4}$ &   $+0.23_{-0.03}^{+0.05}$ &  1.65 &
 $12.4_{ -0.7}^{ +0.5}$ &   $+0.23_{-0.03}^{+0.05}$ &  1.65 &
 $11.6_{ -0.5}^{ +0.4}$ &   $+0.27_{-0.02}^{+0.02}$ &  2.59 &
 $12.0_{ -0.3}^{ +0.1}$ &   $+0.30_{-0.02}^{-0.00}$ &  5.56\\

4478 & $ 7.2_{ -1.3}^{ +1.3}$ &   $+0.22_{-0.07}^{+0.05}$ &  0.12 &
 $ 7.6_{ -1.6}^{ +2.1}$ &   $+0.23_{-0.06}^{+0.04}$ &  0.12 &
 $ 7.7_{ -1.5}^{ +1.6}$ &   $+0.21_{-0.07}^{+0.06}$ &  0.10 &
 $ 8.6_{ -1.3}^{ +1.1}$ &   $+0.27_{-0.03}^{+0.02}$ &  0.03\\

4489 & $ 4.6_{ -0.8}^{ +0.6}$ &   $+0.09_{-0.08}^{+0.08}$ &  2.82 &
 $ 5.6_{ -1.6}^{ +2.3}$ &   $+0.10_{-0.09}^{+0.09}$ &  2.69 &
 $ 5.0_{ -0.9}^{ +0.9}$ &   $+0.06_{-0.07}^{+0.09}$ &  2.84 &
 $ 4.3_{ -0.8}^{ +0.5}$ &   $+0.25_{-0.05}^{+0.03}$ &  3.83\\

4551 & $ 6.1_{ -0.5}^{ +1.0}$ &   $+0.26_{-0.04}^{+0.03}$ &  3.13 &
 $10.0_{ -2.0}^{ +1.2}$ &   $+0.26_{-0.03}^{+0.02}$ &  2.50 &
 $ 7.3_{ -1.4}^{ +2.1}$ &   $+0.26_{-0.05}^{+0.03}$ &  2.53 &
 $ 9.8_{ -2.2}^{ +2.0}$ &   $+0.28_{-0.03}^{+0.01}$ &  2.29\\

4621 & $10.5_{ -0.3}^{ +0.2}$ &   $+0.27_{-0.02}^{+0.02}$ & 10.73 & 
 $10.6_{ -0.3}^{ +0.3}$ &   $+0.27_{-0.02}^{+0.02}$ & 10.73 &
 $10.9_{ -0.2}^{ +0.2}$ &   $+0.27_{-0.02}^{+0.02}$ & 10.67 &
 $12.2_{ -0.5}^{ +0.2}$ &   $+0.30_{-0.02}^{-0.00}$ &  8.56\\

4697 & $ 8.1_{ -0.4}^{ +0.6}$ &   $+0.27_{-0.02}^{+0.02}$ &  1.65 &
 $ 8.1_{ -0.5}^{ +0.7}$ &   $+0.27_{-0.02}^{+0.02}$ &  1.65 &
 $ 8.3_{ -0.6}^{ +1.0}$ &   $+0.26_{-0.04}^{+0.03}$ &  1.80 &
 $11.0_{ -1.3}^{ +1.1}$ &   $+0.27_{-0.03}^{+0.01}$ &  0.24\\
\hline
\end{tabular}
\end{minipage}
\end{center}
\end{table*}
\normalsize

\subsection{Comparison with the side-band method}

Figure ~\ref{fig:EWSBBMC} shows a comparison of the line
strengths measured on the same spectra using the side-band (horizontal
axes) and the BMC methods (vertical axes). Notice the departure from a
simple linear relationship, mainly caused by the different way
neighbouring lines affect the estimate of the pseudo-continuum.  In
figures \ref{fig:Balmer1}~and~\ref{fig:Balmer2} we illustrate in more
detail the difference in the measurement of the equivalent width of
H$\gamma$ and H$\delta$, looking more closely at the effect of nearby
lines on the indices.  In order to eliminate the dependence of the EWs
on velocity dispersion we classify the sample into two subsets,
according to velocity dispersion. Figure~\ref{fig:Balmer1} considers
only those galaxies with $\sigma\leq 150$~km/s and
figure~\ref{fig:Balmer2} focuses on galaxies with $\sigma\leq
200$~km/s. In both cases the galaxies are smoothed to the maximum
velocity dispersion of the subsample, in order to make a consistent
comparison. We note that by comparing spectra smoothed to some maximum
$\sigma$ one would obtain a false sense of agreement between the
methods, since the effect of smoothing removes information from the
spectra.

The left panels of figures
\ref{fig:Balmer1}~and~\ref{fig:Balmer2} compare the EWs of the
galaxies in each subset (open black circles), but we focus on two
galaxies in each case (solid squares).  The galaxies are chosen
because they have a similar value of the EW using one method, and a
significantly different value using the other method. The spectral
regions targeted in those galaxies is shown in the rightmost
panels. The straight short line intersecting the spectra is the
side-band pseudo-continuum and runs between the centres of the
flanking side band regions The continuous line running along the top
of the spectra is the BMC continuum. These are shown in grey and black
for each galaxy, as labelled.  The spectra are normalised to allow us
to overplot both SEDs.  In the case of H$\gamma$ we follow
~\cite{va99} and normalise at the peak of 4365\AA. For H$\delta$ we
normalise to an average flux value of 1 within the wavelengths of the
blue side-band, to highlight the effect of CN absorption towards the
red side of the line.

In figure~\ref{fig:Balmer1} we show those galaxies with
$\sigma\leq 150$~km/s. In the top panels, we focus on NGC 4458
and NGC 4467, since these galaxies give a similar side-band H$\delta$
EW but very different BMC-based measurements.  Looking in detail at
the SEDs on the right, the red side-band
of NGC4467 is more affected by increased absorption, mainly caused by
the CN features which, as reported in \cite{yam06}, are stronger in
NGC4467 (CN$_1$=0.110~mag), than in NGC4458 (CN$_1$=0.074~mag). Notice
that this does not strongly affect the BMC pseudo-continuum.  In
addition, the H$\delta$ absorption observed in NGC 4467 is
possibly underestimated relative to NGC 4458. The iron abundance
reported by \cite{yam06} is higher in NGC 4467 by $\sim$0.1~dex, an
effect that would be hidden by the normalisation over the iron lines
around $\sim$4070\AA.  In the bottom panels of
figure~\ref{fig:Balmer1} we investigate the H$\gamma$ lines in NGC
4387 and NGC 4467 which have a similar EW using the BMC method,
whereas they differ by $\sim$0.5~\AA\ in the side-band method.  The
SEDs reveal that these two galaxies do indeed have similar Balmer
absorption, but the side-band method is 'tricked' when comparing the
flux with the continuum (straight line). The increased G4300
absorption in NGC 4387 pushes the pseudo-continuum down, which reduces
the EW value incorrectly. This is consistent with the measured 
absorption at this resolution of NGC4467 (G4300$_{20}$ = 6.5~\AA)
compared to that in NGC4387 (G4300$_{20}$ = 7.2~\AA).

Shown in figure~\ref{fig:Balmer2} is the subsample with
higher velocity dispersion, 
comparing galaxies at $\sigma\leq 200$km/s. The top panel
shows the considerable discrepancy between the side band and BMC
methods on measuring H$\delta$. Since the over-abundance of CN and the
resulting absorption tends to increase with the mass of the galaxy
\citep[see e.g. ][] {sbla03,tol09}, the inclusion of the more massive
NGC 4473 serves to show how significant the effect can be. The
$\sim$0.6\AA\  difference in H$\delta$ given by the side band method can
be seen to be generated by the increase in absorption in the red
side-band region $\sim$4120\AA, of NGC4473. We note again that the CN
absorption, identified by \cite{yam06} is higher in NGC 4473
(CN$_1$=0.138~mag), than in NGC 4478 (CN$_1$=0.077~mag).  The bottom
panels display a more subtle situation in which the effect of the
surrounding metal lines is to reverse the relative values of the
H$\gamma$ EWs.  The increased absorption, again mainly coming from the
G4300 feature, forces the pseudo-continuum to a lower value, resulting in a 
small EW. The effect, while small, serves to indicate how
sensitive the side band method can be to the surrounding spectral features.

\begin{figure}
\begin{center}
\includegraphics[width=3.2in]{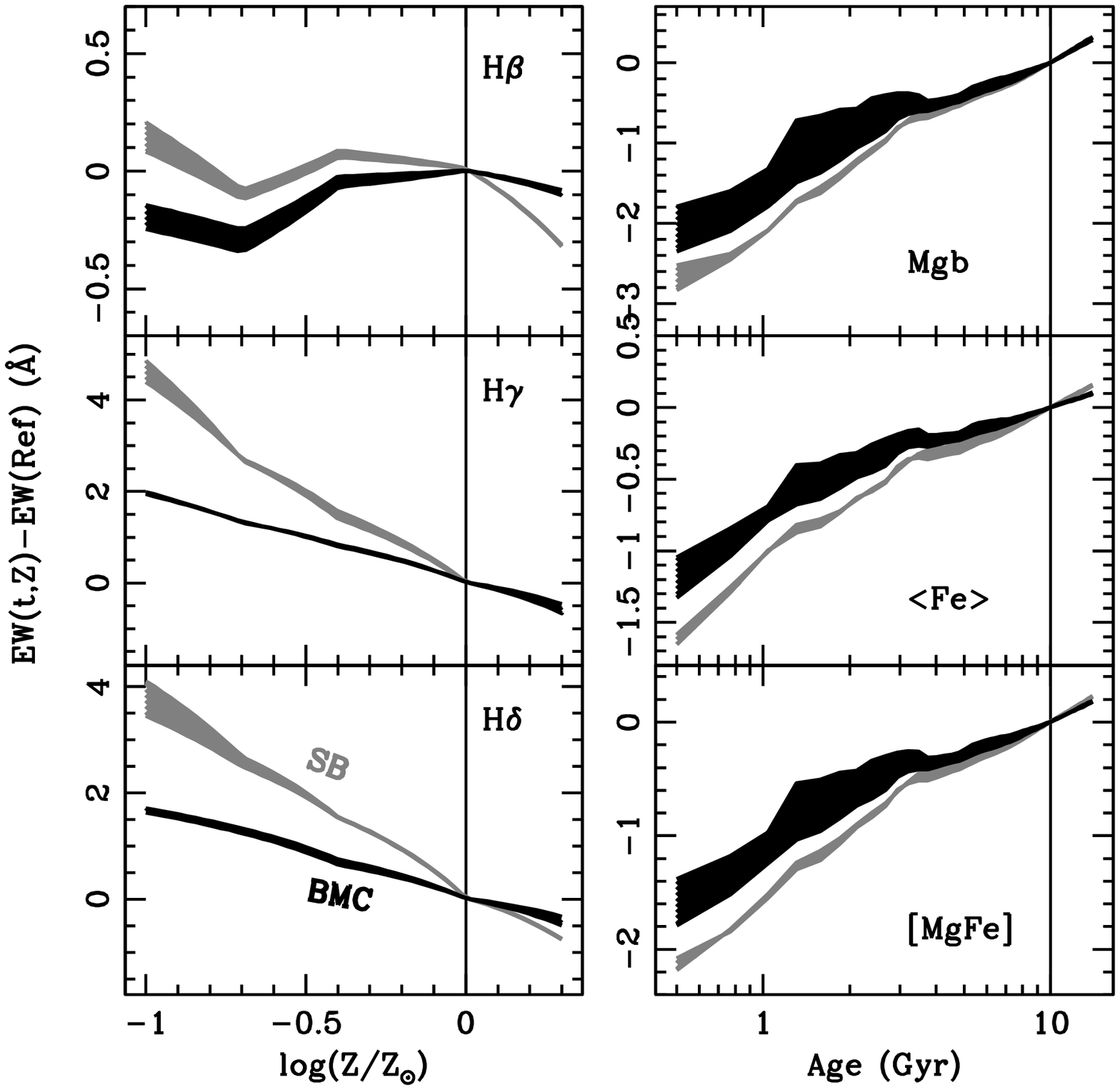}
\caption{Dependence of Balmer lines on metallicity (\emph{left})
and ``metal'' lines on age (\emph{right}). The standard side-band
method (SB; grey) and our proposed Boosted Median Continuum (BMC; black)
are shown for a range of metallicities and ages as shown. On the
left, the shaded areas correspond to an age range $[6,12]$~Gyr. On the
right, the shaded areas span a range of metallicities: $-0.3<\log Z/Z_\odot <+0.3$.
The vertical axis is the difference between the equivalent width of the
line for a given metallicity (left) or age (right) and the value at
a reference point given by the vertical line (i.e. solar metallicity 
on the left and 10 Gyr on the right).}
\label{fig:Idx}
\end{center}
\end{figure}

Therefore, figures~\ref{fig:Balmer1}~and~\ref{fig:Balmer2} show that 
our proposed method is more resilient to the effects of neighbouring 
absorption lines. To further illustrate this point, figure~\ref{fig:Idx} 
compares the age-metallicity degeneracy of typical Balmer (\emph{left}) and metal
lines (\emph{right}) measured either in the standard way or using the
BMC pseudo-continuum. The shaded areas indicate the difference in the
EWs with respect to a reference stellar population: on the left it is
the population with the same age at solar metallicity, and on the
right the reference is a population with the same metallicity and
10~Gyr old, both references marked by the vertical lines.  Black and
grey shading correspond to the BMC and the side-band methods,
respectively. The panels on the left explore the
age-sensitive Balmer indices for a range of metallicities and the
panels on the right show metal indices for a range of ages. Ideally, a
perfect observable would result in a horizontal line (i.e. zero
metallicity dependence of an age-sensitive line and vice-versa). The
models span a wide range of ages and metallicities as shown in the
caption. This figure shows that BMC-based measurements of EWs are less
subject to the age-metallicity degeneracy than the side-band
methods. This result is especially dramatic for H$\gamma$ and
H$\delta$, for which the age-metallicity degeneracy drops from
$\Delta$EW$/\Delta\log(Z/Z_\odot)= -4.2$ (side-band) to $-1.9$ (BMC)
in H$\gamma$ or from $-3.9$ (side-band) to $-1.7$ (BMC) in H$\delta$
(values measured at the fiducial 10~Gyr, solar metallicity SSP).
Furthermore, the shaded regions of the metal-line indices on the
rightmost panels are much wider for the BMC method (black), showing
that at a fixed age, the BMC method spans a wider range of EW, thereby
being more sensitive to metallicity (see caption for details).

\begin{figure}
\begin{center}
\includegraphics[width=3.2in]{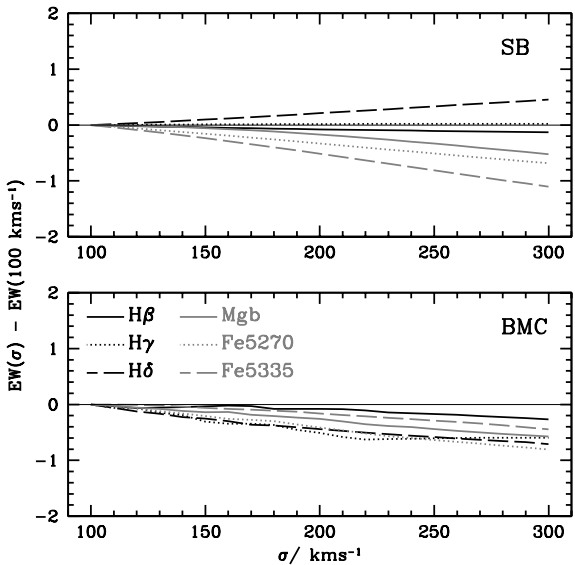}
\caption{The uncertainties caused by an error in the
estimate of velocity dispersion are illustrated by a 
comparison of the difference between the EW measured
at a fiducial value ($\sigma=$100km/s) and a range
of velocity dispersions (horizontal axis). Various
line strengths are considered, as labelled, for the
standard SB method ({\sl top}) and our proposed BMC
method ({\sl bottom}).}
\label{fig:sigerr}
\end{center}
\end{figure}

In order to quantify the effect of an uncertainty in the velocity
dispersion on the EWs, we compare in figure~\ref{fig:sigerr} the
change in EW between a fiducial value ($\sigma=$100km/s) and a range
of velocity dispersions (horizontal axis). The SB (BMC) values are
shown in the top (bottom) panels for a number of absorption lines, as
labelled. One can see that the effect on BMC-measured EWs can be
corrected much in the same as with the standard SB method. In
addition, for most cases the correction is smaller for BMC estimates.

Furthermore, we have compared the effect of noise on our proposed BMC
pseudo-continuum and found that for a wide range of signal-to-noise
ratios (from 10 to 100 per \AA\ ) the uncertainty in the EW of all
lines is always $\sim 0.3$~dex smaller than those obtained with the
side-band method. 
However due to the lower dynamical range, as the 
BMC is forced to stay above the spectra, the S/N requirements are 
similar to other methods.
We follow \citet{optoHB} and define the S/N at which the method can 
distinguish $\pm$2.5 Gyr at 10 Gyr. For $H\gamma$ and $H\delta$ 
we get a S/N of $\sim 50$, and for $H\beta$ the required S/N is $\sim 60$.

\begin{figure}
\begin{center}
\includegraphics[width=3.2in]{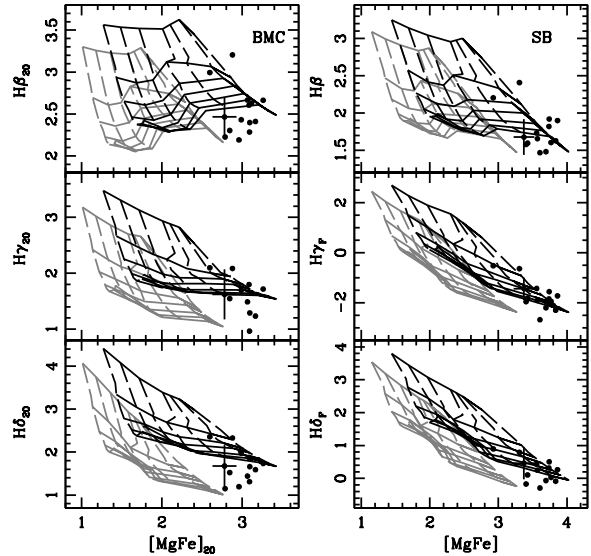}
\caption{Comparison of grids of SSP models using the standard
side-band method (SB, \emph{right}) and our proposed Boosted Median
Continuum (BMC, \emph{left}). A systematic error bar is included in
each panel (the observational error bars are much smaller). We show
the grids for two different velocity dispersions 100km/s (black) and
300km/s (grey). The solid lines connect the SSPs with ages (from top
to bottom) 2 to 14 Gyr in steps of 2 Gyr.  The dashed lines connect
the SSPs with metallicities (from left to right) [m/H]=$-$1 to $+$0.2 in
steps of 0.2~dex.  }
\label{fig:grid}
\end{center}
\end{figure}

Figure~\ref{fig:grid} shows the equivalent widths of several lines for
a grid of SSPs corresponding to a velocity dispersion of 100 km/s
(black) or 300 km/s (grey). The measured EWs (without any correction
for velocity dispersion) are shown as black dots, along with 
a characteristic error bar which mostly comes from the systematics (the
observed uncertainties are much smaller). We emphasize that our
fitting method generates grids corresponding to the measured velocity
dispersion for each galaxy. One can see that BMC EWs (left panels)
appear less degenerate than the grids using a side-band method (right
panels).

\section{Modelling the SFH of elliptical galaxies}

The properties of the unresolved stellar populations of our sample are
constrained by comparing the targeted equivalent widths with four sets
of generic models that describe the star formation history in terms of
a reduced number of parameters. It is our goal to assess the
consistency of different sets of models in fitting \emph{independently}
the different spectral lines targeted as well as the full SED. The
majority of studies in the literature \citep[see
e.g. ][]{kd98,sct00,cald03,tm05} have compared measurements of EWs
with simple stellar populations (i.e. a single age and
metallicity). While those models are probably valid for the
populations found in globular clusters, it is imperative to go beyond
simple stellar populations in galaxies, whose star formation histories
generate complex distributions of age and metallicity.  \cite{fyi04}
and \citet{pas05} showed that composite models of stellar populations
could result in significant differences on the average ages and
metallicities of galaxies. More recently, \citet{sbla06} and
\citet{serra07} have explored this issue through the comparison of two
age indicators, both concluding that composite populations are needed
to consistently model the populations in early-type galaxies.

\begin{figure*}
\begin{minipage}{18cm}
\includegraphics[width=3.2in]{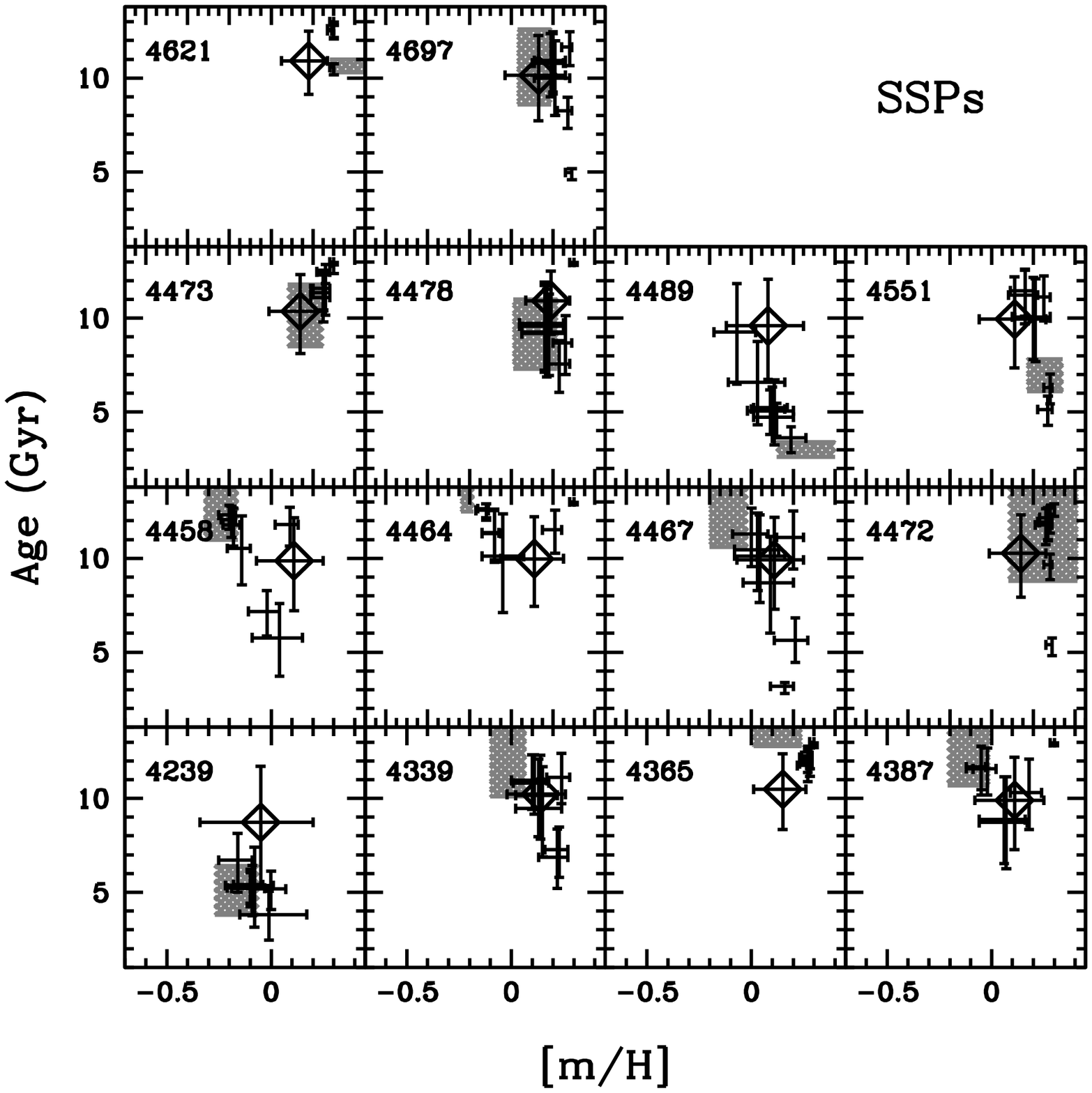}
\includegraphics[width=3.2in]{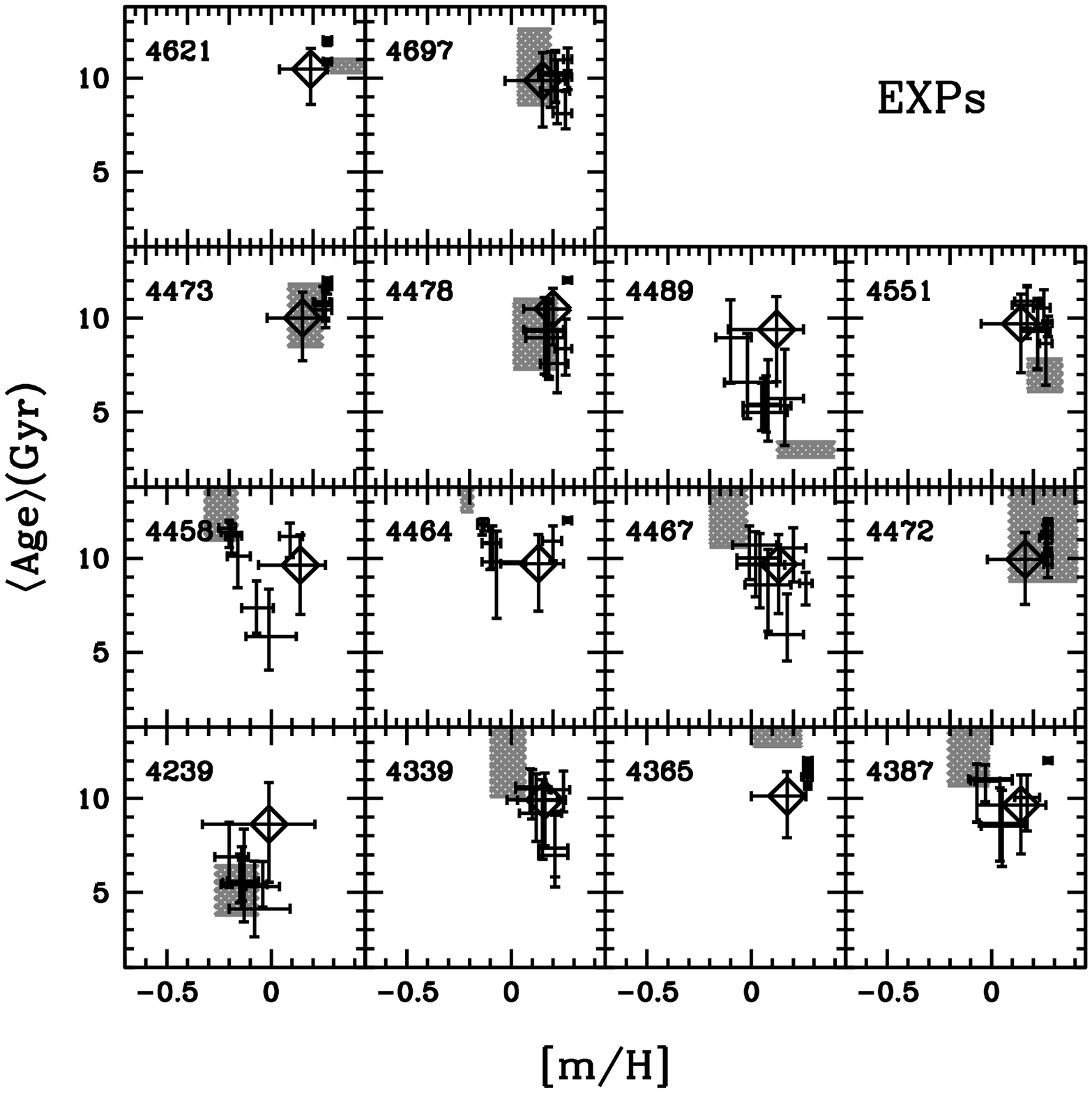}
\caption{Best fit age and metallicity values for our sample,
using SSPs (\emph{left}) or EXP models (\emph{right}). The
error bars shown the 68\% confidence levels and the
shaded regions give the age and metallicity estimates
of \citet{yam06} using the H$\gamma_\sigma$ vs. [MgFe]
diagram. The diamond gives the fit to the age and
metallicity using \emph{only} the spectral
energy distribution (i.e. no line strengths are used for
this data point).}
\label{fig:maps1}
\end{minipage}
\end{figure*}

\begin{figure*}
\begin{minipage}{18cm}
\includegraphics[width=3.2in]{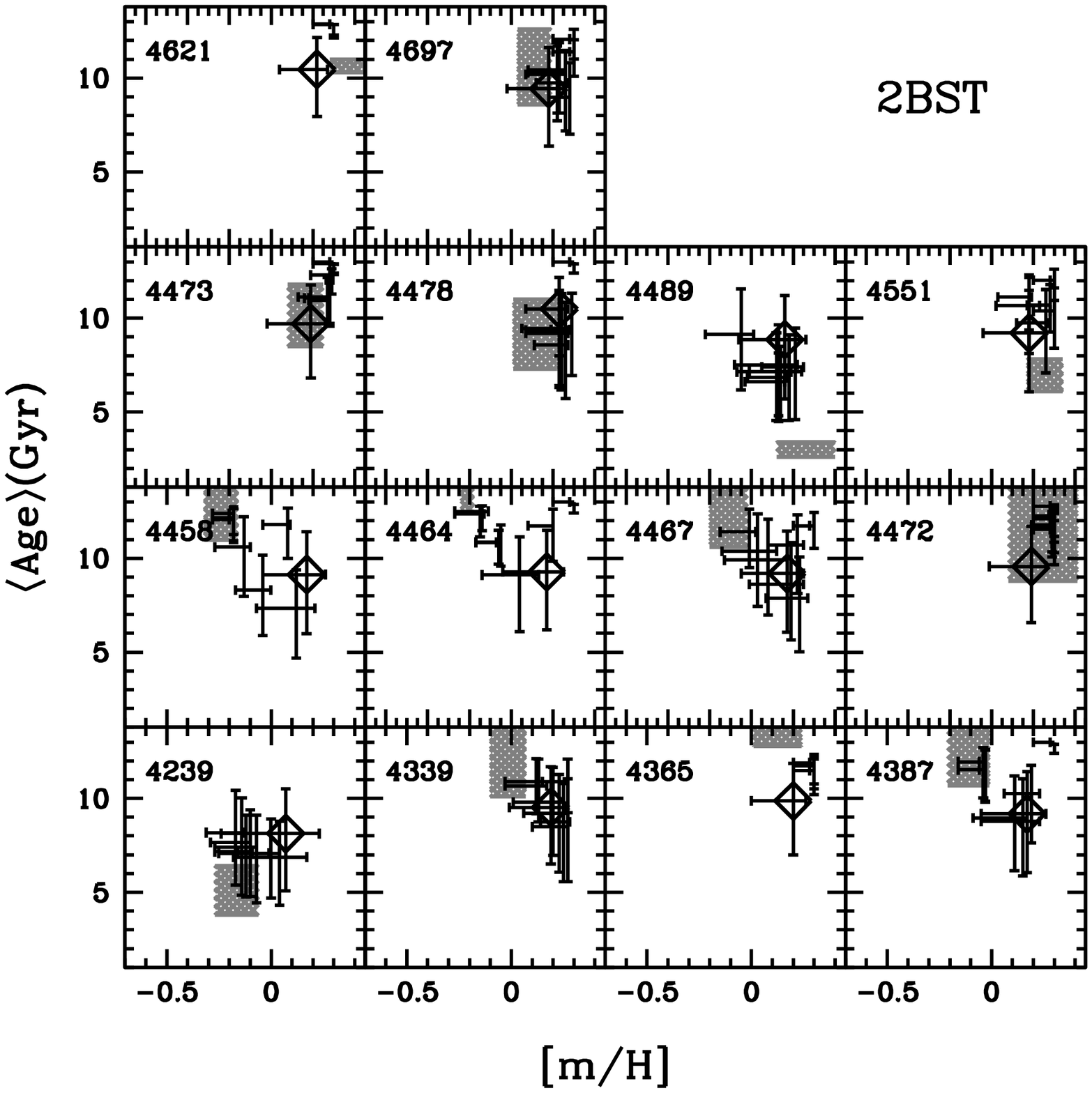}
\includegraphics[width=3.2in]{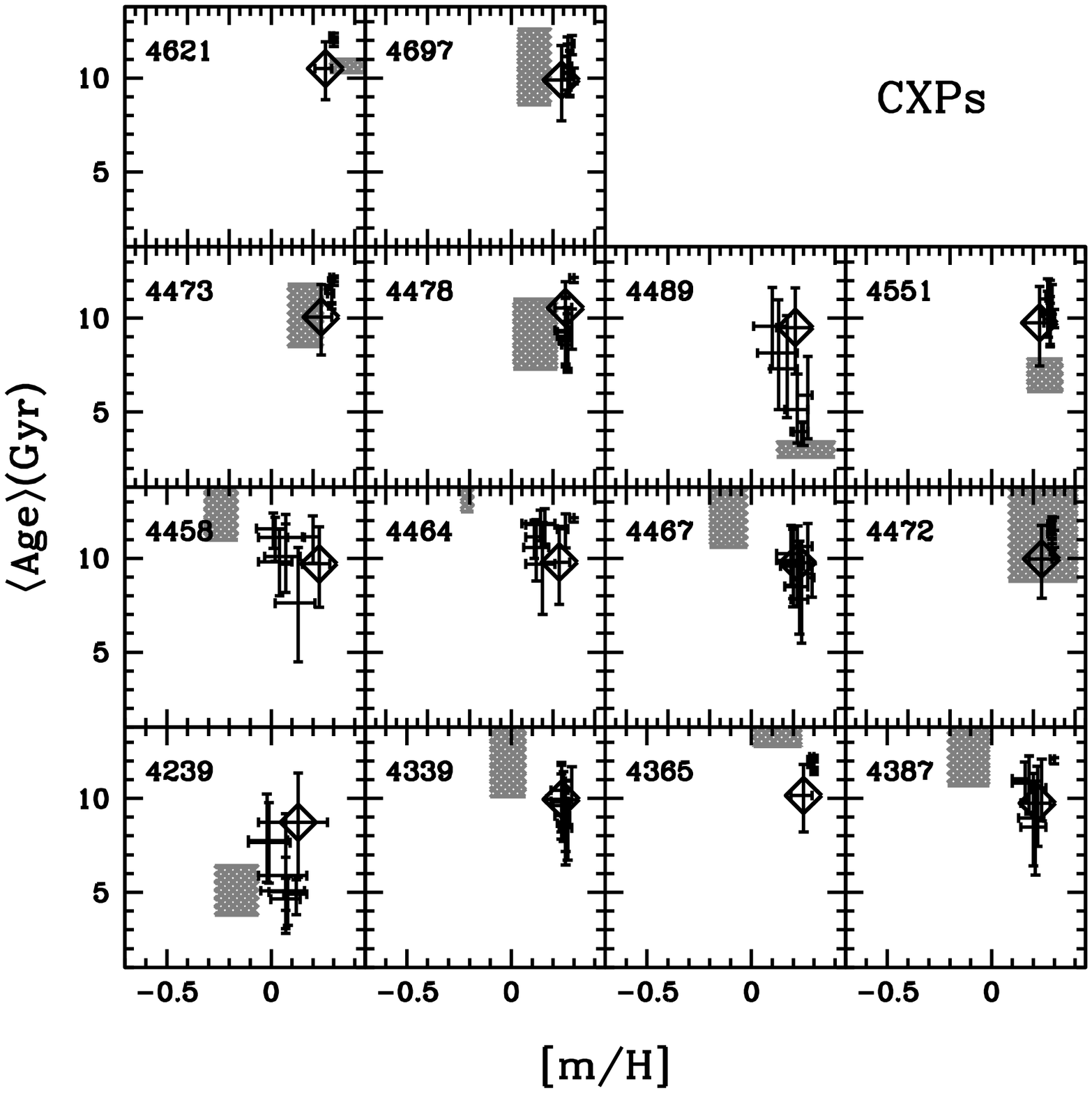}
\caption{Best fit age and metallicity values for our sample,
using 2-Burst (\emph{left}) or CXP models (\emph{right}). The
error bars shown the 68\% confidence levels and the
shaded regions give the age and metallicity estimates
of \citet{yam06} using the H$\gamma_\sigma$ vs. [MgFe]
diagram. The diamond gives the fit to the age and
metallicity using \emph{only} the spectral
energy distribution (i.e. no line strengths are used for
this data point).}
\label{fig:maps2}
\end{minipage}
\end{figure*}

In this paper, except for the first case (namely Simple Stellar
Populations), the models generate a distribution of ages and/or
metallicities that are used to combine the population synthesis models
of \citet{bc03}, assuming a \citet{chab03} Initial Mass Function. We
use the updated 2007 version \citep{cb07}. The resulting synthetic
spectral energy distribution is smoothed to the same resolution and
velocity dispersion of the galaxy. A correction for Galactic reddening
is applied, assuming the \citet{Fitz99} law (this is mostly done for
the comparison of the full SEDs, as EWs are not affected). Finally,
the EWs are computed and compared with the observations using a
standard maximum likelihood method.  The high S/N of the observed
spectra imply that our error budget is mainly dominated by the
systematics of the population synthesis models.  It is not obvious how
to incorporate problems such as sparsely populated stellar parameter
space or systematics within the models, into the analysis.  As a
compromise we estimate the uncertainties associated with the models by
generating 500 Monte Carlo simulations of gaussian noise at the level
of S/N $\sim$50, that of the stellar library at the core of the
BC03/CB07 models \citep[STELIB][]{stelib}.  These are added in quadrature
with those generated for each galaxy index.  

\begin{figure}
\begin{center}
\includegraphics[width=3.2in]{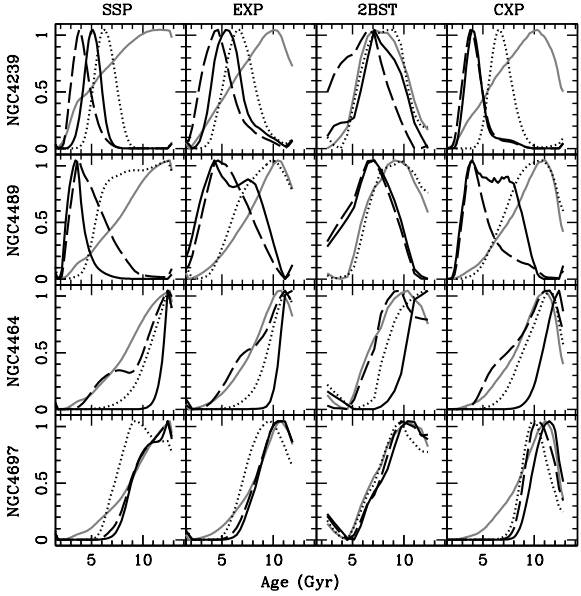}
\caption{Likelihood distribution of the marginalised average stellar ages for some
of the Virgo cluster elliptical galaxies from our sample. The
distribution is shown for the four models as labelled. The grey solid 
line corresponds to the fit of the SED. The lines correspond 
to the age distribution according to [MgFe] plus either H$\beta$ (black,solid), 
H$\gamma$ (black,dashed) or H$\delta$ (black,dotted),
respectively.}
\label{fig:AgeHist}
\end{center}
\end{figure}

The four sets of models considered in this paper are listed below,
with the range of parameters shown in table~\ref{tab:params}.

\begin{enumerate}
\item {Simple Stellar Populations (SSP):} This has been the most
popular method used in the analysis of the ages and metallicities of
early-type galaxies. The advantage lies in it simplicity: SSPs are the
building blocks of all population synthesis models, and they are
carefully calibrated against realistic SSPs (i.e. globular
clusters). An SSP is primarily defined by an age and a metallicity
although it is also dependent on the abundance pattern and the IMF.
The drawback of this method is that the parameters obtained are
inherently \emph{luminosity weighted}, such that a small amount of
young stars can have a significant effect on the age extracted with
this method. Furthermore, SSPs are not expected to model galaxy
populations, which have an extended range of ages and
metallicities. By using SSPs to model early-type galaxies one makes
the assumption that the stellar populations have a narrow age
distribution compared to stellar evolution timescales.
 
\item {Exponential SFH (EXP):} 
A more physical scenario should consider an extended period of star
formation. The EXP models (also called $\tau$ models in the
literature) model the star formation rate as an exponentially
decaying function of time with timescale $\tau$, started at an epoch
given by a formation redshift $z_F$. The metallicity is assumed to be
fixed throughout the SFH and is also left as a free parameter.

\item {Two Burst Formation (2BST):} 
Recent rest-frame NUV observations have revealed the presence of
residual star formation in early-type galaxies \citep[see
e.g.][]{sct00,fs00,yi05,kav07}.  The presence of small amounts of
young stars can significantly affect the derived SSP model parameters
\citep{serra07}.  This model describes this mechanism with two simple
stellar populations.  Four parameters are left free: the age of the
old (t$_O$) and the young components (t$_Y$), the mass fraction in
young stars (f$_Y$), and the metallicity of the populations ($Z$,
assumed to be the same for the old and the young components).

\item {Chemically Enriched Exponential (CXP):} 
All the models considered above fix the metallicity throughout the
SFH. More realistically, a model should incorporate in a consistent
way the buildup of metallicity caused by previous generations of
stars. These chemical enrichment models \citep[see e.g.][]{bp99,fs03}
include the stellar yields from intermediate and massive stars and
result in a distribution of metallicities as well as ages. The aim of
this paper is to explore simple models that can be easily
implemented. Hence, instead of applying a detailed model of chemical
enrichment, we define in a purely phenomenological way a model that
mimics those. We keep the same SFH as in the EXP models (described by
a star formation timescale $\tau_1$ and a formation redshift
$z_F$). Furthermore, the metallicity is assumed to increase in step with 
the cumulative distribution of the star formation rate, namely:

\be
Z(t) = Z_1 + (Z_2 - Z_1) \Big[ 1 - \exp( - \Delta t/\tau_2)\Big],
\ee

\noindent
where $\Delta t = t-t(z_F)$ and $\tau_2$ is the timescale
corresponding to the buildup of metallicity. As a first-order
approach it is valid to assume that metallicity increases with the
stellar mass of the system (roughly, the cumulative distribution of
the star formation rate). More accurately, this timescale will depend
on the star formation efficiency or on the fraction of gas ejected in
outflows \citep[see e.g. ][]{fs03}. The upper and lower values of
metallicity are fixed to $Z_1=Z_\odot /10$ and $Z_2=2Z_\odot$,
although the average metallicity will be controlled by the timescale
$\tau_2$.
\end{enumerate}

\begin{figure}
\begin{center}
\includegraphics[width=3.2in]{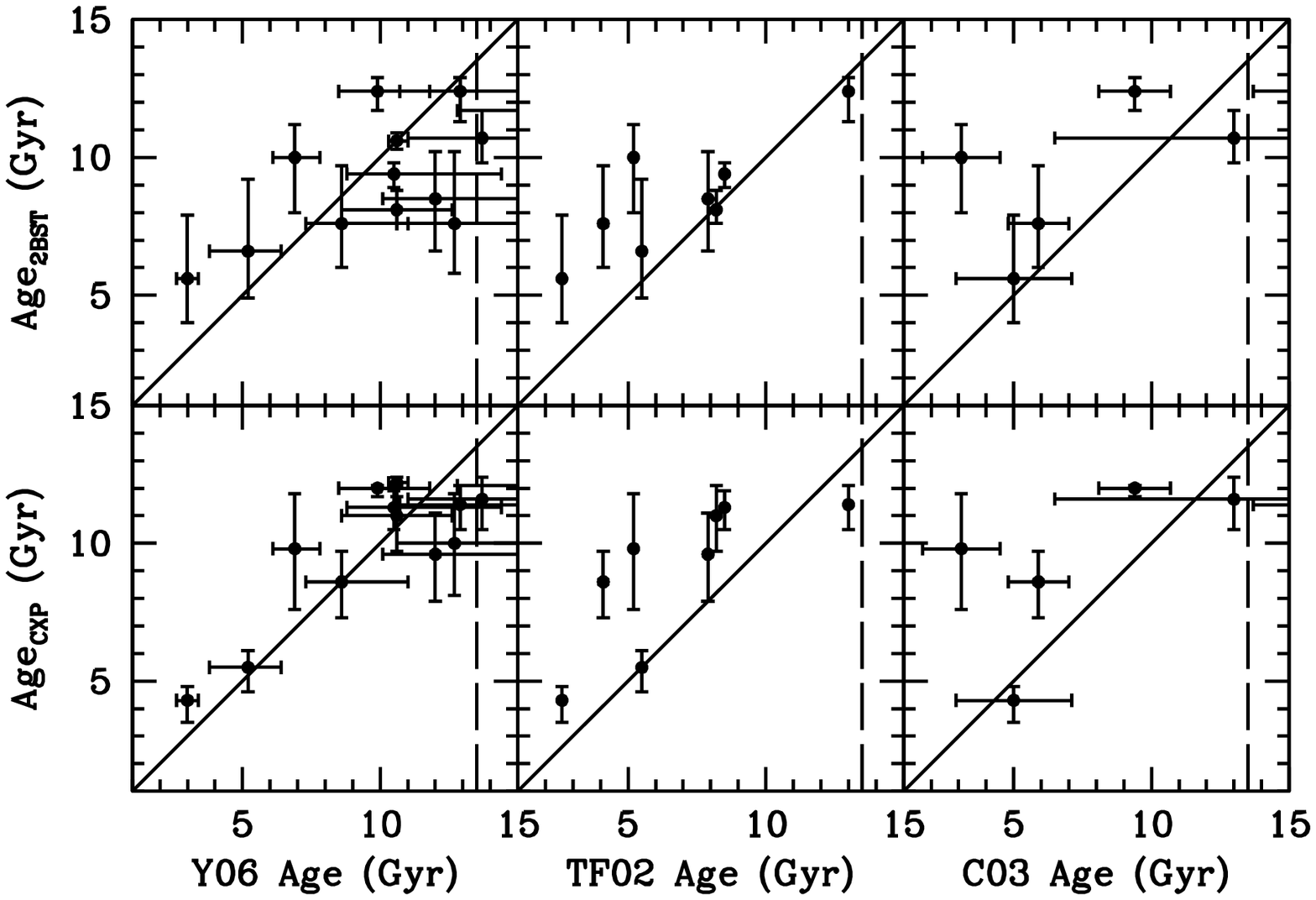}
\caption{Comparison between the ages presented here and those from
previous studies. Y06, TF02 and C03 indicate \citep{yam06},
\citep{tf02}, and \citep{cald03}, respectively. The figure illustrates
the difference in the average ages when including two types of
composite populations: a 2-burst model (2BST; top) or a smooth star formation
history including chemical enrichment (CXP; bottom).
The slanted solid line
is a 1:1 correspondence.  The vertical dashed line represents the age
of the Universe for our cosmology, and is used as a prior in the
analysis (see text for details).}
\label{fig:Ages}
\end{center}
\end{figure}

We assume a standard $\Lambda$CDM cosmology ($\Omega_m$=0.3; H$_0$= 70
km/s/Mpc) to constrain the maximum ages of the stellar populations to
the age of the Universe (13.7~Gyr). One could allow the fitting
algorithm to stray into older ages to explore wider volumes of
parameter space, but this paper takes the cosmological constraints on
the age of the Universe at face value \citep{wmap}. Furthermore, population
synthesis models are poorly calibrated for populations older than
Galactic globular clusters. We note that previous work on the same
spectra \citep{yam06} probed the population synthesis models of
\citet{va99} all the way to their oldest available ages ($\sim$20~Gyr),
using the isochrones of \citet{bert94}.
Although in only two cases, NGC4365 at 20~Gyr and NGC4464 at 18.5~Gyr, 
did they find cosmologically contradictory ages, we still choose to avoid 
this by limiting the maximum age.

\begin{table*}
\caption{Best fit parameters for the 2BST and CXP models}
\label{tab:BFit}
\begin{tabular}{lrllllll}
\hline\hline
Galaxy &  M$_{\rm s}^1$ & \multicolumn{3}{c}{2BST} & \multicolumn{3}{c}{CXP}\\
NGC & $\times 10^{10}$M$_\odot$ & t$_O$(Gyr) & t$_Y$(Gyr) & f$_Y$ & $\tau_1$(Gyr) & $\tau_2$(Gyr) & z$_F$\\
\hline
4239 &  0.9(0.7)& $ 7.9_{ -2.7}^{ +2.7}$ &   $ 2.6_{ -0.3}^{ +0.3}$ &   $0.08_{-0.06}^{+0.12}$ 
& $0.2_{ -0.2}^{ +0.5}$ & $0.2_{ -0.1}^{ +0.5}$ & $0.6_{ -0.1}^{ +0.2}$\\
4339 &  4.6(4.6)&  $ 8.8_{ -2.0}^{ +1.8}$ &   $ 1.8_{ -1.3}^{ +1.1}$ & $\simlt 0.01$
& $0.4_{ -0.2}^{ +0.4}$ & $0.0_{ -0.0}^{ +0.1}$ & $1.7_{ -0.5}^{ +0.8}$\\
4365 & 38.0(36.3)&$11.9_{ -0.9}^{ +0.7}$ &   $ 1.5_{ -1.1}^{ +1.0}$ & $\simlt 0.01$
& $0.8_{ -0.2}^{ +0.2}$ & $2.9_{ -2.0}^{ +2.0}$ & $3.0_{ -0.2}^{ +0.2}$\\
4387 &  3.2(2.8)& $12.4_{ -0.8}^{ +0.5}$ &   $ 1.6_{ -1.1}^{ +1.0}$ & $\simlt 0.01$
& $0.7_{ -0.3}^{ +0.4}$ & $0.3_{ -0.1}^{ +0.2}$ & $2.6_{ -0.5}^{ +0.5}$\\
4458 &  3.1(2.2)&$10.9_{ -1.0}^{ +0.7}$ &   $ 1.6_{ -1.1}^{ +1.0}$ & $\simlt 0.02$
& $0.2_{ -0.1}^{ +0.4}$ & $0.2_{ -0.2}^{ +0.6}$ & $2.5_{ -0.4}^{ +0.6}$\\
4464 &  1.8(1.4)& $12.7_{ -0.5}^{ +0.2}$ &   $ 1.6_{ -1.1}^{ +1.0}$ & $\simlt 0.01$
& $0.6_{ -0.2}^{ +0.3}$ & $0.4_{ -0.2}^{ +0.2}$ & $2.8_{ -0.4}^{ +0.3}$\\
4467 &  0.4(0.4)& $ 8.6_{ -2.6}^{ +2.6}$ &   $ 2.4_{ -1.1}^{ +0.5}$ &   $0.03_{-0.03}^{+0.11}$
& $0.2_{ -0.1}^{ +0.1}$ & $0.1_{ -0.1}^{ +0.1}$ & $1.7_{ -0.6}^{ +0.8}$\\
4472 & 97.7(102.3)& $ 9.5_{ -0.8}^{ +0.4}$ &   $ 1.6_{ -1.1}^{ +1.0}$ & $\simlt 0.02$
& $0.9_{ -0.4}^{ +0.6}$ & $0.2_{ -0.2}^{ +0.2}$ & $2.8_{ -0.4}^{ +0.3}$\\
4473 & 20.4(20.4)& $12.5_{ -0.8}^{ +0.4}$ &   $ 1.5_{ -1.1}^{ +1.0}$ & $\simlt 0.01$
& $1.0_{ -0.1}^{ +0.2}$ & $1.6_{ -1.1}^{ +1.1}$ & $3.0_{ -0.2}^{ +0.2}$\\
4478 &  4.7(5.6)& $ 8.3_{ -2.2}^{ +2.1}$ &   $ 1.9_{ -1.3}^{ +1.0}$ &  $0.02_{-0.02}^{+0.05}$ 
& $0.6_{ -0.3}^{ +0.5}$ & $0.1_{ -0.1}^{ +0.1}$ & $1.4_{ -0.3}^{ +0.6}$\\
4489 &  1.2(1.2)&$ 6.9_{ -2.5}^{ +2.4}$ &   $ 2.6_{ -0.4}^{ +0.3}$ &   $0.13_{-0.10}^{+0.18}$ 
& $0.3_{ -0.1}^{ +0.3}$ & $0.1_{ -0.1}^{ +0.1}$ & $0.4_{ -0.1}^{ +0.1}$\\
4551 &  4.0(2.8)& $10.7_{ -2.0}^{ +1.7}$ &   $ 2.7_{ -0.2}^{ +0.2}$ &  $0.04_{-0.03}^{+0.06}$ 
& $0.3_{ -0.1}^{ +1.1}$ & $0.2_{ -0.1}^{ +0.1}$ & $1.7_{ -0.6}^{ +0.9}$\\
4621 & 33.9(33.1)& $10.7_{ -0.4}^{ +0.1}$ &   $ 1.5_{ -1.1}^{ +1.0}$ & $\simlt 0.02$
& $1.0_{ -0.3}^{ +0.3}$ & $1.4_{ -1.0}^{ +1.0}$ & $4.1_{ -1.2}^{ +0.9}$\\
4697 & 17.0(12.3)&$ 8.3_{ -0.7}^{ +0.4}$ &   $ 1.6_{ -1.2}^{ +1.0}$ &  $\simlt 0.03$
& $0.3_{ -0.1}^{ +0.4}$ & $0.2_{ -0.1}^{ +0.1}$ & $2.3_{ -0.4}^{ +0.5}$\\
\hline
\end{tabular}

$^1$ Stellar masses are computed from the best fit CXP (2BST)
models. Estimated uncertainty $\Delta\log$M$_s\sim 0.3$~dex (mostly
from the assumption about the IMF).
\end{table*}

\subsection{Disentangling the stellar populations}

The comparison of the EWs is done by measuring the line strengths
directly from the model SEDs after being resampled and smoothed to the
resolution and velocity dispersion of the actual observations of each
galaxy, taken from \citet{yam06}. Since the spectra of the models have
already been smoothed in terms of velocity dispersion to that of the
real galaxy, no correction for this need be applied. All galaxies are
fitted using the age-sensitive indices, H$\beta_{20}$, H$\gamma_{20}$,
H$\delta_{20}$ and G4300$_{20}$ along with the metallicity
indicator [MgFe]$_{20}$\citep{gon93}. This is used instead of the
newer [MgFe]$^\prime$ as defined by \citet{tm03} since the original is
almost independent of $\alpha/$Fe and it is not clear whether the same
fractions of the two Fe lines is still appropriate for the new BMC
indices. We note that we have excluded the D4000 index 
since it was found to be poorly modelled by BC03, a problem already 
identified by both \citet{wild07} and \citet{sbla06c}. The exclusion (or inclusion) of 
the D4000, does not significantly affect the final results obtained although 
it does significantly increase (decrease) the quality of the fit.
The removal of the D4000 also limits the constraints we can put on 
second order parameters (e.g. age and mass fraction of young stellar 
population in the 2-Burst model).

As indicated by \citet{schiavon04} and later confirmed by
\citet{serra07}, a discrepancy found between the parameters derived
through each of the three Balmer lines could be explained by the
existence of an underlying younger sub-population. \citet{schiavon04}
explain that this may well be due to the dominance of the younger
population at bluer wavelengths. However the mere fact that different
indices would give different parameters, should suggest that the model
being used has not captured all aspects of the true star formation
history. Hence, in the analysis of EWs of our sample galaxies, we are
not only looking to determine the age and metallicity but also
estimating whether it is possible to \emph{rule out a single
population} to describe an early-type galaxy. This could be done
through discrepancies between the three Balmer lines targeted
here, a possibility highlighted by \citet{schiavon07} on a single
stacked spectra of low mass ellipticals, who found that the addition
of a small young population alleviated such discrepancies.

With the same aim, we focus on possible discrepancies between 
individual line strengths and explore a number of star formation
histories to determine the age distribution of our sample of elliptical
galaxies.

\subsection{Spectral Fitting}
As an alternative approach to targeted line strengths, we also
consider the full spectral energy distribution to constrain
the star formation history of our sample. In this paper
we want to test the consistency between the constraints imposed
by the line strengths and those from a full spectral fit.

We use a maximum likelihood method to fit the spectra, which is
analysed over the spectral range between 3900\AA\ and 4500\AA\ (or
4000--4500\AA\ for the six galaxies observed with WHT). The SEDs are
normalised over the same range. The data span a much wider spectral
range (out to 5500-5800\AA ).  However, we have chosen a smaller
range to avoid flux calibration errors, which could introduce
important systematic changes in the predicted ages and metallicities. 
The spectral range chosen straddles the D4000 break, since this is a
very sensitive indicator of the ages of the stellar populations.

\subsection{Results}

Figures~\ref{fig:maps1} and \ref{fig:maps2} show the average age and
metallicity for all four models explored in this paper. The error bars
are shown at the 69\% confidence level (i.e. like a 1$\sigma$ level
although our analysis does not assume a Gaussian distribution). 
For the SSPs the average age is trivially the age of the
population. For the others, the average values are weighed by the
stellar mass. Each error bar corresponds to the \emph{individual fit}
of an age-sensitive line (H$\beta$, H$\gamma$,
H$\delta$ or G4300) plus the [MgFe] as metallicity indicator
(i.e. each point comes from the analysis involving 2 indices). The
diamond and its error bar correspond to the fit to the SED -- no
additional information from the line strengths is added to this
likelihood. In grey, the shaded areas correspond to the ages from
\citet{yam06} (using the H$\gamma_\sigma$ -- [MgFe] diagram on
SSPs). One can see that SSPs (left panel of figure~\ref{fig:maps1})
show the largest discrepancies between the ages and metallicities
estimated using independently the different age-sensitive
indices. This would imply that a joint likelihood putting all indices
together would not be a consistent way of determining the ages of the
populations.

Nevertheless, as a comparison we show in table~\ref{tab:Ages} the
best estimates of the average age and metallicity, combining all five
indices (but not the fit of the SED). Table~\ref{tab:BFit} shows the
best fit parameters for the 2BST and CXP models. The uncertaintes in
both tables are given at the 90\% confidence level.  It can be
seen from table~\ref{tab:Ages} that in a majority of cases the use of
a model with an age spread offers an improvement in the $\chi^2$ value
for the five indices.  Importantly the recovered values of metallicity
remain consistent (within errors) across the single metallicity models
considered, indicating its robustness against the inclusion of
additional populations. For some galaxies, such as NGC 4365, NGC 4387
and NGC 4473, an SSP is as good as the other models presented here and
in fact the parameters recovered by complex SFHs essentially reduce to
an SSP. This is to be expected since the short formation timescale
expected for elliptical galaxies will in some cases be well
represented by an SSP. However galaxies which have younger SSP ages
give very different (mass-weighted) ages when considering a 2-Burst
model.  One interesting result is that the CXP models, which
incorporate a spread both in metallicity and age, find solutions
that are as good as -- and in the case of NGC4458, and NGC4467 and NGC
4697 significantly better than -- the other models.

In addition, the discrepancies among individual measurements shown in
figure~\ref{fig:maps1} reflects the fact that a single age and
metallicity scenario is a weaker, less consistent model of the
populations of a galaxy. Table~\ref{tab:Ages} shows that the actual
change in the age when going beyond SSPs can be quite significant,
especially for younger ages. We emphasize here that the different ages
do not reflect a possible inclusion of a prior caused by the choice of
parameters. All the composite models presented here (CSP, EXP and
2-Burst) explore a range of parameters that includes the best-fit ages
and metallicities obtained by the SSPs.

In some cases the $\chi^2$ values are larger than might be
expected. In the case of NGC~4621, the largest contributor is the
G-band. If we remove that indicator from the analysis, the reduced
$\chi^2$ decreases to $\sim 1$.  It is clear that the high S/N of the
spectra poses a challenge for current stellar population synthesis
models. Certainly the errors associated with the models do not take
into consideration internal problems with the models themselves or the
possible limitations of the stellar libraries from which they are
assembled. \cite{proct04} also consider that limitations in the models
as the cause of the high $\chi^2$ values.  

\begin{figure}
\begin{center}
\includegraphics[width=3.2in]{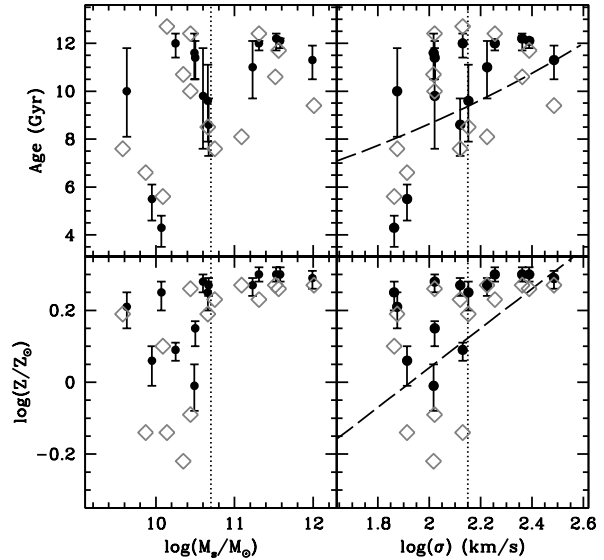}
\caption{Ages and metallicities versus stellar mass (left)
and velocity dispersion (right). The solid dots (grey diamonds)
correspond to the estimates from the CXP (2-Burst) models.
The dashed lines in the rightmost
panels are the age and metallicity scaling relation of \citet{tm05}.}
\label{fig:AgeMet}
\end{center}
\end{figure}

The mismatch discussed above can be seen in more detail in figure~\ref{fig:AgeHist},
where the marginalised distribution of average age for all four models is shown
for a few galaxies from the sample, as labelled. The black solid, dashed, and dotted lines
correspond to individual fits to H$\beta$, H$\gamma$, and H$\delta$, respectively,
along with the G4300 (grey, dotted) and the SED (grey, solid). SSPs (leftmost panels) fail
to give a consistent distribution, whereas any of the other models
give a more unified distribution. Notice that the galaxies shown in
the top panels (NGC 4239 and NGC 4489) have young populations, whereas
the bottom two galaxies (NGC 4464 and NGC4697) have overall older populations.
The top two galaxies are better fit by a 2-Burst model and the bottom two get
better fits from an extended model like CXP. This would suggest that the youngest
populations in early-type galaxies are best fit by the assumption of 
``sprinkles'' of young stars, as suggested by \citet{sct00} and modelled by 
\citet{fs00} and \citet{kav07}.

This explains why the models incorporating an age spread have
greater maximum likelihood values, where $\mathcal{L}_{max} \propto
\exp(-\chi^2_{min}/2)$.  However, as was seen in table~\ref{tab:Ages},
the fits of the SSPs are not substantially worse than the more complex
models. This is not surprising given firstly, that the SFH of many
elliptical galaxies is not different from a single episode of star
formation, especially when it happens at high redshift. Secondly, as
identified by \cite{serra07}, for bursts of mass fraction, $f_{\rm Y}
\geq$ 10\% the disagreement in the recovered ages from different
Balmer lines dissappears.  This indicates that there is no
differential effect across the spectrum from which to determine an age
spread. Nevertheless, it remains the case that the more complex models
are physically well motivated, resulting in better average age
estimates and fit the data more consistently. Under such criteria we
select the 2BST and CXP for further analysis.

Figure~\ref{fig:Ages} shows a comparison of our age estimates using
2-Burst (\emph{top}) and CXP models (\emph{bottom}) with values taken
from the literature, as given in the figure caption. It is important
to notice that our models do not consider ages older than the age of
the Universe using a concordance $\Lambda$CDM cosmology (i.e. 13.7
Gyr). Our ages are also closer to a mass weighted average due to
the underlying assumptions in the modelling in both cases. We find
that for the youngest galaxies, our analysis gives rather older ages
than previous estimates from the literature. Also notice that the
older galaxies appear younger because of the strong constraint imposed
by the chosen cosmological parameters. Figure~\ref{fig:AgeMet} shows
the average age and metallicity as a function of stellar mass (\emph{
left}) and velocity dispersion (\emph{right}), both for the CXP (black
dots) and for the 2-Burst models (grey diamonds). The stellar masses
are obtained by combining the absolute magnitude of the galaxies in
the $V$ band with $M/L_V$ corresponding to the best fit CXP models
(There is no significant difference if the best fit 2-Burst models are
used instead). The dashed line is the fit to age and metallicity from
\citet{tm05}. Our age-mass relationship is compatible with recent
estimates in the literature \citep[see
e.g. ][]{sct00,cald03,tm05,sbla06b}. We should emphasize that our
reduced sample of Virgo \emph{elliptical} galaxies excludes massive,
young S0s in high density environments \citep{tm05}, removing them as
a possible cause of the age scatter.

\begin{figure}
\begin{center}
\includegraphics[width=3.2in]{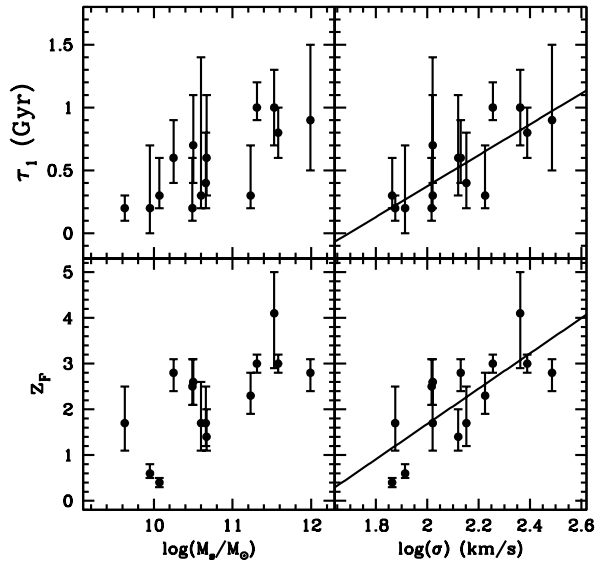}
\caption{The predictions of the CXP models for star formation timescale (\emph{
top}) and formation redshift (\emph{bottom}) are shown with respect to
stellar mass (\emph{left}) and velocity dispersion (\emph{right}). 
This speculative plot suggests that the age trend found in
figure~\ref{fig:AgeMet} is mainly caused by a range of formation epochs (bottom).
The formation timescale (top) also correlates with mass or velocity dispersion,
but it does not have values larger than $\tau_1\sim 1.5$~Guyr.
The solid lines in the
right panels are simple linear fits to the data (equations~\ref{eq:zF}~\&~\ref{eq:Tau},
see text for details).}
\label{fig:zF}
\end{center}
\end{figure}

Furthermore, notice in figure~\ref{fig:AgeMet} the transition in the age
distribution around 150 km/s in velocity dispersion or
5$\times$10$^{10}M_\odot$ in stellar mass. Above this value the ages
of the populations are overall old and with small scatter among
galaxies, whereas lower mass galaxies have a wide range of ages and
metallicities (not necessarily younger throughout).  The trend is
robustly independent of the modelling, as shown both by the CXP fits
(black dots) and the 2-Burst models (grey diamonds). This threshold is
reminiscent of the one found by \citet{kauf03} in the general
population of SDSS galaxies at a stellar mass $3\times 10^{10}M_\odot$
or the threshold in star formation efficiency of late-type galaxies at
rotation velocities v$_c\sim$140 km/s based on their photometry
\citep{tfz} or on the presence of dust lanes \citep{dal04}.

\begin{figure*}
\begin{minipage}{15.5cm}
\begin{center}
\includegraphics[width=5.2in]{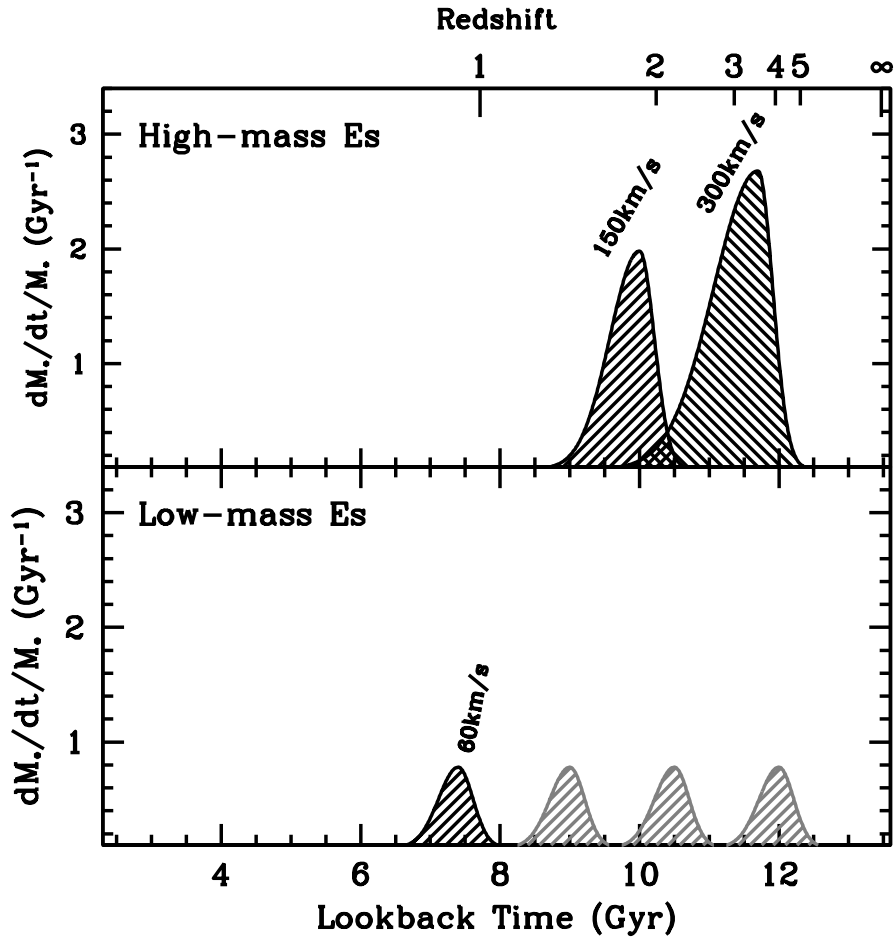}
\caption{Our proposed model for the star formation history 
of Virgo cluster elliptical galaxies. The parameter estimates suggest
a nearly constant star formation timescale with respect to velocity dispersion,
whereas formation \emph{epoch} is strongly correlated. We find a stronger
correlation with velocity dispersion (compared to stellar mass), so we label
the formation scenarios according to $\sigma$. The bottom panel 
emphasizes the issue that the \emph{observed} low-mass Es are best fit
by later formation epochs, an issue which does not rule out the possibility
of other low-mass Es in the past -- or elsewhere -- to have been formed 
earlier (grey Gaussians).}
\label{fig:SFH}
\end{center}
\end{minipage}
\end{figure*}

\section{Conclusions}

We have revisited the superb, high signal-to-noise spectra of 14
elliptical galaxies in the Virgo cluster presented by
\citet{yam06,yam08}. Our main goal was two-fold: a) to give an optimal
but versatile definition of the equivalent width of spectral features
that would minimise the contribution from nearby lines, b) to explore 
the possibility of discriminating between the standard treatment of
galaxy spectra either as simple stellar populations or composite models 
with a distribution of ages and metallicities.

The former was achieved by defining a ``boosted median'' pseudo-continuum.
This method is very easy to implement on any spectral data and it
improves on the traditional side-band methods by reducing the final
uncertainty of the EWs for the same spectra and reducing the
age-metallicity degeneracy of age- and metal-sensitive lines by
reducing the effect of unwanted, nearby spectral features.
The method only requires two numbers to determine the pseudo-continuum
(i.e. confidence level of the boosted median and size of the kernel
over which the analysis is performed). We propose 90\% and 100\AA\
for these two parameters when dealing with stellar populations of
galaxies at moderate resolution (R$\sim 1000-2000$).

The second goal -- going beyond SSPs -- is approached by comparing
SSPs and three more sets of models which assume some distribution of
ages and metallicities. We find simple populations fail to give a
consistent marginalised distribution of ages when fitting
independently different age-sensitive lines. We propose either a
2-Burst model or a $\tau$ star formatiom history with a
phenomenological prescription for chemical enrichment (CSP
models). They give similar results, with a clear trend between average
age and either stellar mass or velocity dispersion, as shown in
figure~\ref{fig:AgeMet}.

As suggested above when discussing figure~\ref{fig:AgeHist}, if
younger (older) galaxies are better fit by 2-Burst (CXP) models, one
would expect that young stars in elliptical galaxies appear in a
random way, and not as a time-coherent, smooth distribution. This
result supports minor merging -- possibly involving metal-poor
sub-systems -- as the main mechanism to generate recent
star formation in early-type galaxies \citep{kav09}. This process
will be readily detected in low-mass galaxies, whereas a similar
amount of young stars in a massive galaxy will be harder to detect.

\begin{figure*}
\begin{minipage}{18cm}
\includegraphics[width=3.2in]{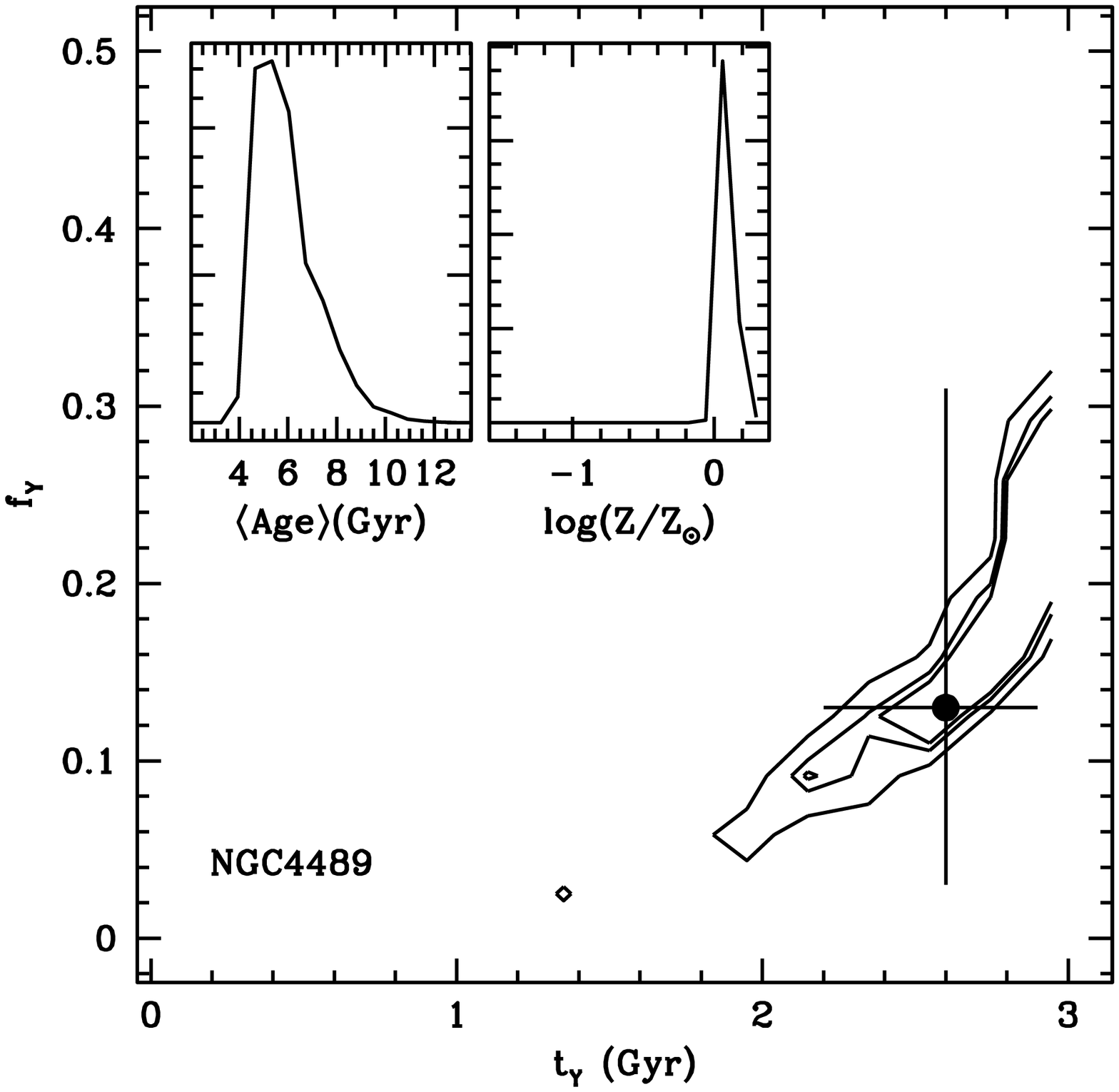}
\includegraphics[width=3.2in]{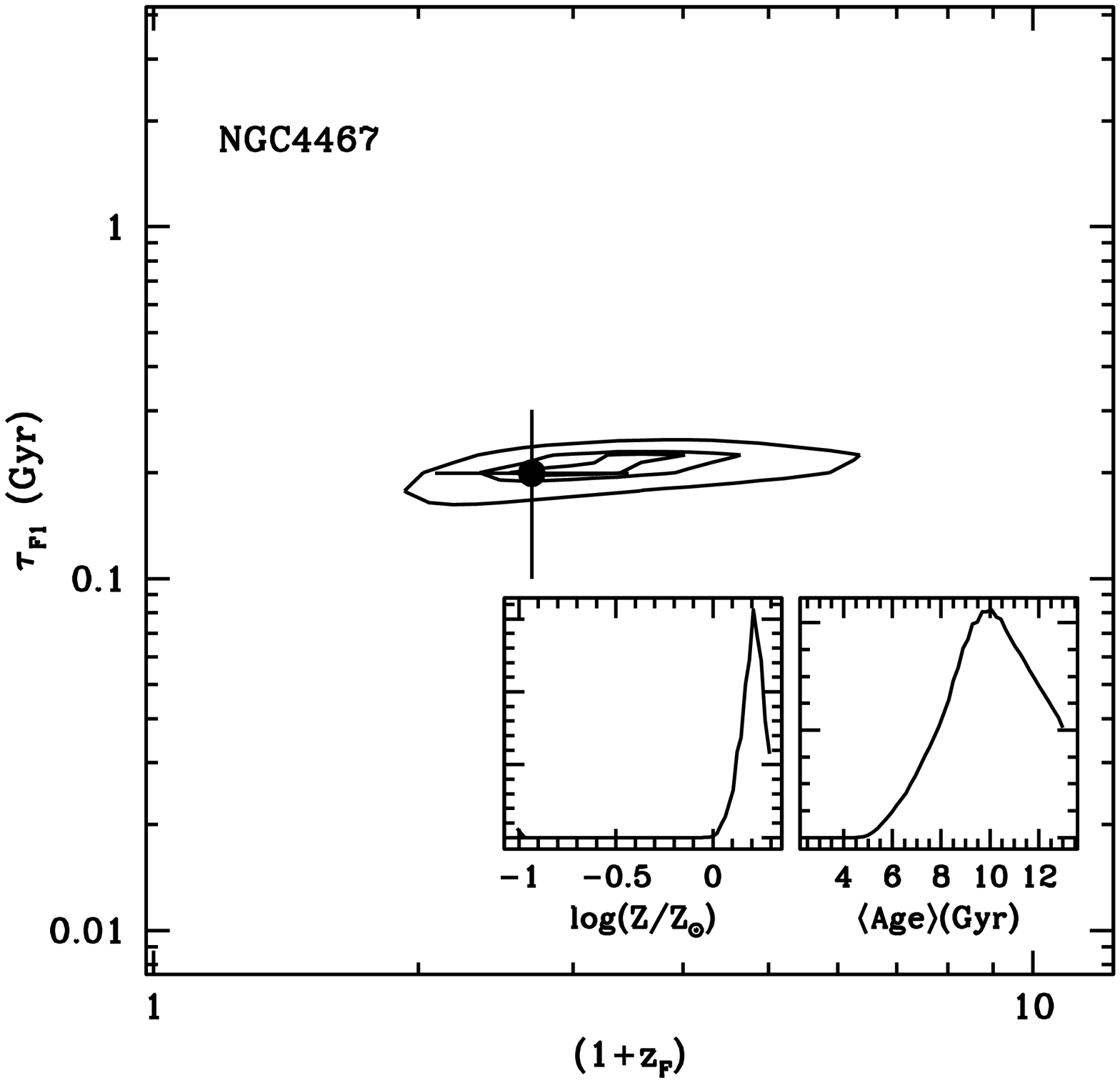}
\caption{Probability distributions of NGC 4489 (left) and NGC 4467
(right) for 2-Burst and CXP models, respectively.  \emph{Left:} The
contour plot shows the degeneracy between the age (t$_{\rm Y}$) and
mass fraction (f$_{\rm Y}$) of the younger component. Contours are
shown at the 75, 90 and 95\% confidence level.  The solid circle
identifies the best fit and quoted error bars.  The inset shows the
probability distributions of the average age and metallicity, both of
which are well constrained.  \emph{Right Panel} The degeneracy between
the formation redshift (z$_{\rm F}$) and star formation timescale
($\tau_1$) is shown in the contour plot (75,90 and 95\% confidence
levels).The probability distribution of the average age and
metallicity (inset) show that their values are well constrained.}
\label{fig:Degen}
\end{minipage}
\end{figure*}

Doubtlessly average ages and metallicities are the observables best
constrained by these models. Parameters like formation timescale and
formation epoch can only be considered ``next-order'' observables,
which will be fraught with larger uncertainties.  This is largely
due to the massive degeneracies which are present in such parameters
as mentioned in the introduction. Its is well known that the mass and
age of a subpopulation are degenerate in that additional mass fraction
can compensate for an older age or vice versa \citep{leo96}. Another
considerable degeneracy also exists between the formation redshift and
the star formation timescale for both the EXP and CXP models
\citep[see e.g. ][]{gobat08}.  While our indices cover a large
wavelength range we find that these degeneracies are still
significant. Figure~\ref{fig:Degen} shows the degeneracy between
parameters for a typical case using 2-Burst models (\emph{left}) and
CXP models (\emph{right}). We take the probability weighted values
across the entire parameter space as a more robust measure of these
parameters.  Although the derived parameters are still subject to
considerable uncertainty, as expressed by the error bars shown in
figure~\ref{fig:zF}, we note that our grids sample the parameter space
quite finely meaning our recovered likelihood distribution should be a
reasonable approximation to the complete distribution.

We proceed in a more speculative way, taking the continuous CXP
models at face value and putting some trust on the predictions of their
formation epochs and timescales. Figure~\ref{fig:zF} reveals an
intriguing trend that suggests it is not just formation timescale but
also formation epoch which drives the mass-age relationship in early-type
galaxies. The solid lines in the top/bottom right panels gives a simple
linear fit to the data, namely:
\be
z_F \sim  1.8 + 3.9\log\sigma_{100}
\label{eq:zF}
\ee
\be
\tau ({\rm Gyr}) \sim  0.3 - 1.14\log\sigma_{100}
\label{eq:Tau}
\ee

\noindent
with $\sigma_{100}$ given in units of 100~km/s. Figure~\ref{fig:SFH}
shows our proposed model for the star formation history of Virgo
cluster elliptical galaxies. We model the formation histories as
skewed Gaussians whose spread maps the star formation timescale, and
we include a correlation between formation epoch and velocity
dispersion, as labelled. Notice that the \emph{observed} low-mass
galaxies (black in the bottom panel) are best fit by late formation
epochs, although this fact does not prevent smaller early-type
galaxies in the past to have formed at earlier epochs (represented
by the grey Gaussians). Our sample is too small and too local to
explore this issue further.

One could argue that the lack of a correlation between formation
timescale ($\tau_1$; \emph{top}) and mass would contradict the observed
relationship between mass and abundance ratios \citep[see
e.g. ][]{tm05}. One possibility would involve minor merging of dwarf
galaxies. Those mergers will have a more prominent effect on the
spectra of low-mass galaxies. Dwarf galaxies have a very extended
period of star formation and eject a big fraction of their metals,
resulting in metal-poor gas with solar or sub-solar abundance ratios
that will reduce the observed (luminosity-weighted) [$\alpha$/Fe] in
low-mass galaxies. Alternatively, notice that the formation epoch
($z_F$; \emph{bottom}) correlates quite strongly with velocity
dispersion, implying that the structures that form the bulk of the
stars in low-mass galaxies are more likely to be contaminated by the
ejecta from type~Ia supernova, thereby reducing the abundance ratios
to reproduce the observed correlation between [$\alpha$/Fe] and
mass. A more detailed analysis of the abundance ratios -- although
beyond the scope of this paper -- is under way to explore this very
interesting possibility.

\section*{Acknowledgments}
We would like to thank the referee, Ricardo Schiavon, for his very
insightful and extensive comments and suggestions. This work has made
use of the delos computer cluster in the physics department at King's
College London.


\end{document}